\documentclass[%
reprint,
 amsmath,amssymb,
 aps,
showkeys
]{revtex4-2}
\usepackage{tikz}
\usetikzlibrary{quantikz2}
\tikzset{%
    uctrl/.style={draw, circle, minimum size=0.285pc, append after command={ \pgfextra { \fill  (0,0)-- (225:2pt) arc (225:405:2pt) -- cycle; } } }
}
\def\myvdots{\ \vdots\ }
\DeclareExpandableDocumentCommand{\uctrl}{O{}{m}}{|[uctrl,#1]| {#2} \qw}

\usepackage[dvipsnames]{xcolor}
\newif\ifproofread
\newcommand{\changemarker}[1]{%
\ifproofread
\textcolor{red}{#1}%
\else
#1%
\fi
}

\usepackage{dsfont}
\usepackage{adjustbox}
\usepackage{braket}
\usepackage{babel}
\usepackage{physics}
\usepackage{tabularx}
\newcolumntype{Y}{>{\centering\arraybackslash}X}
\usepackage{siunitx}
\usepackage[utf8]{inputenc}
\usepackage{nicematrix}% Nice matrices
\usepackage{graphicx}% Include figure files
\usepackage{dcolumn}% Align table columns on decimal point
\usepackage{bm}% bold math
\usepackage{booktabs}
\usepackage{float}
\usepackage{hyperref}% add hypertext capabilities

\begin{document}
\proofreadfalse
\preprint{APS/123-QED}

\title{Quantum Simulation-Based Optimization for Cooling System Design}
\author{Leonhard H\"olscher$^{1,2,3}$}
\email[Corresponding author: ]{leo.h@mail.de}
\author{Lukas Karch$^{1}$}
\author{Or Samimi$^{4}$}
\author{Tamuz Danzig$^{4}$}

\affiliation{$^{1}$BMW Group, 80809 Munich, Germany}
\affiliation{$^{2}$Theoretical Physics, Saarland University, 66123 Saarbr\"ucken, Germany}
\affiliation{$^{3}$Institute for Quantum Computing Analytics (PGI 12), Forschungszentrum J\"ulich, 52425 J\"ulich, Germany}
\affiliation{$^{4}$Classiq Technologies, 6473104 Tel Aviv-Yafo, Israel}
\date{\today}% It is always \today, today,

\begin{abstract}
Engineering design processes involve iterative design evaluations requiring numerous computationally intensive numerical simulations. Quantum algorithms promise substantial speedups for specific tasks relevant to engineering simulations. However, these advantages quickly vanish when considering data input and output on quantum computers. The recently introduced Quantum Simulation-Based Optimization (QuSO) framework circumvents this limitation by treating simulations as subproblems within a larger optimization problem. Here we adapt and implement QuSO for a simplified cooling system design problem, validate correctness in statevector simulations, and present a detailed gate-level complexity analysis for a single QuSO iteration. We express the scaling in terms of problem parameters and QAOA depth and iterations. We show that the cost function can be coherently computed over a superposition of exponentially many configurations using circuits of polynomial complexity. This does not yield a speedup for a single simulation instance, but it enables potential advantages arising from the subsequent QAOA-based search over configurations. The study serves as a proof-of-concept for integrating fault-tolerant quantum subroutines with simulation-based optimization in engineering workflows, clarifying both promise and practical limitations.
\end{abstract}
\maketitle

%\tableofcontents

\section{Introduction}
Numerical simulations of physical phenomena are central to science and engineering. As models become more detailed and accurate, the associated computational cost grows accordingly. Quantum computers are known for their potential to outperform classical computers in applications ranging from machine learning to optimization problems \cite{riofrio_quantum_2026, dalzell_quantum_2025}. Consequently, the question arises of whether quantum computers can also benefit the field of computer-aided engineering.

Numerous research papers on solving differential equations using quantum computers have appeared in recent years. These quantum approaches are based on analog quantum simulation \cite{jin_analog_2024}, variational quantum algorithms (VQAs) \cite{kyriienko_solving_2021, lubasch_variational_2020, jaksch_variational_2023, demirdjian_variational_2022}, and fault-tolerant quantum algorithms \cite{berry_high-order_2014, berry_quantum_2017, montanaro_quantum_2016}. The latter, in particular, allow a detailed algorithmic complexity analysis, revealing under what conditions a quantum advantage exists. These fault-tolerant quantum algorithms often rely on quantum linear system solvers \cite{harrow_quantum_2009, childs_quantum_2017, gilyen_quantum_2019, martyn_grand_2021, costa_optimal_2022, dalzell_shortcut_2024, morales_quantum_2025} since solving a linear system of equations is a core step of many numerical simulation methods, such as the finite element method \cite{axelsson_finite_2001, landau_survey_2008}. 

However, the anticipated speedup of these quantum algorithms is often diminished by the algorithmic overhead associated with data input and output \cite{aaronson_read_2015}. The recently proposed framework, Quantum Simulation-Based Optimization (QuSO), attempts to circumvent this limitation \cite{stein_exponential_2024} by focusing not on the simulation results, but rather on optimized parameters derived from simulations. This approach is known as simulation-based optimization \cite{wang_simulation_2013, nguyen_review_2014}. The optimization within QuSO can be efficiently addressed by well-known quantum optimization algorithms such as QAOA \cite{farhi_quantum_2014, hadfield_quantum_2019}. Instead of constructing a cost Hamiltonian, QuSO implicitly evaluates the cost function through simulations, which typically involve solving at least one linear system of equations. Next to quantum linear system solvers and quantum optimization algorithms, this approach leverages quantum algorithms for block-encoding of matrices \cite{dalzell_quantum_2025, sunderhauf_block-encoding_2024, camps_explicit_2024, childs_hamiltonian_2012} and amplitude estimation \cite{lomonaco_quantum_2002}. 

Here we adapt QuSO to solve a simplified cooling system design problem in which the decision variables specify thermal connections in a resistive network. Despite our simplification of the cooling system, it is inspired by a real-world use case and thus of industrial relevance. A quantum algorithm for a similar problem has been studied by Sato et al. \cite{sato_quantum_2023}. Their objective is to find the optimal ground structure based on thermal material properties. Therefore, they also construct a VQA for the optimization task, however, they employ another VQA to solve the underlying linear system. While the authors argue that their algorithm benefits from the simultaneous evaluation of all possible ground structures, the full variational design of their algorithm does not allow an explicit statement about its gate complexity.

Since our implementation uses fault-tolerant subroutines, we can provide a transparent gate-level complexity analysis for a single QuSO iteration. We show that we coherently compute the temperature as an objective function in superposition for exponentially many configurations using a polynomial number of gates. While this does not yield a speedup for a single simulation, or solving one linear system of equations, it enables a potential advantage arising from QAOA on top. We complement this analysis with statevector simulations that validate correctness and show depth–precision trade-offs on the simplified cooling model. Overall, this work serves as a proof-of-concept and illustrates an end-to-end implementation of QuSO rather than a performance benchmark.

The remainder of the paper is organized as follows. Section~\ref{sec:problem} introduces the cooling system model and its formulation as a linear system of equations. In Section~\ref{sec:QuSO}, we detail our adaptation of QuSO, including all quantum subroutines and their algorithmic complexities, to encode the required matrix, solve the linear system, and transfer the result to the optimization algorithm. Section~\ref{sec:simulations} presents the results of our numerical simulations and analysis. Finally, we discuss the algorithm and potential improvements in Section~\ref{sec:discussion}, while Section~\ref{sec:conclusion} summarizes our findings.

\section{Problem Formulation}\label{sec:problem}
A vehicle is composed of numerous complex components, ranging from mechanical systems like the drivetrain to electrical systems, such as batteries, and electric motors. These systems are interconnected, influencing one another, and must meet a wide range of requirements. For instance, depending on weather conditions, the battery may require heating or cooling while still providing sufficient power to the motor. Therefore, the components are connected by a cooling system with a cooling liquid, ensuring that all components have their desired temperature. 

\begin{figure}[h!]
    \centering
    \includegraphics[]{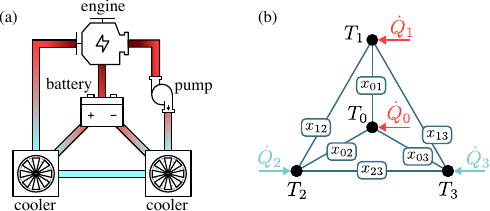}
    \caption{Cooling system model. (a) Example cooling system with an engine and battery as heat sources, two coolers, and a pump. (b) The simplification of the same cooling system that only covers heat conduction. $x_{ij}$ is a binary variable defining if a connection exists between two nodes, $T_i$ is the temperature at each node, and $\dot{Q}_i$ is the heat transfer rate from either a heat or cooling source.}
    \label{fig:problem}
\end{figure}

In this paper, we focus on simulating the vehicle's cooling system, as illustrated in Fig.\,\ref{fig:problem}\,(a). 

\noindent We apply several simplifications to define an initial quantum algorithm to address the problem. Heat transfer generally occurs through convection, conduction, and radiation \cite{lienhard_heat_2024}. Given that convection is a highly nonlinear phenomenon, and quantum operations are fundamentally linear, we omit this effect in our current approach for the sake of simplicity. While certain nonlinear problems can be addressed using linearization techniques on quantum computers \cite{lloyd_quantum_2020, liu_efficient_2021, jin_quantum_2024}, this lies beyond the scope of this work. Moreover, we do not consider the effects of radiation for now. This allows us to model the cooling system without a pump since we do not consider the motion of the cooling liquid. Instead, a system with $N$ components or nodes is defined by thermal resistances $R_{ij}$ of every connection between nodes $i$ and $j$, and every heat or cooling source is defined by its external heat transfer rate $\dot{Q}_i$, as depicted in Fig.\,\ref{fig:problem}\,(b) for $N=4$. In addition, we have binary variables within
\begin{equation}
    \mathbf{x} = \{x_{ij}\}_{(i, j)\in\mathbf{E}}
\end{equation}
defining if a connection between the two respective nodes exists. $\mathbf{E}$ defines the set of $m$ edges that we consider in our graph. If the graph is fully connected, we have
\begin{equation}
    \mathbf{E} = \{(i, j) | 0 \leq i < j \leq N-1\}\mathrm{.}
\end{equation}

Analogous to Ohm's law, Fourier's law of heat conduction in its simplest form is given by \cite{lienhard_heat_2024}
\begin{equation}
    \dot{Q} = \frac{\Delta T}{R_\mathrm{th}}\mathrm{,}
    \label{eq:ohm}
\end{equation}
where $\Delta T$ is the temperature difference between two nodes, and $R_\mathrm{th}$ is the corresponding thermal resistance. Energy conservation requires that the heat entering a node equals the heat leaving it. This principle in combination with Eq.\,\eqref{eq:ohm} leads to a linear system of equations
\begin{equation}
    \mathbf{A}(\mathbf{x})\mathbf{\tilde{T}} = \mathbf{B}\mathrm{,}
    \label{eq:ls}
\end{equation}
where $\mathbf{A}(\mathbf{x})\in\mathbb{R}^{N\times N}$, $\tilde{\mathbf{T}}\in\mathbb{R}^{N}$, and $\mathbf{B}\in\mathbb{R}^{N}$. The solution vector contains the temperatures at each node
\begin{equation}
    \mathbf{\tilde{T}} = (\tilde{T}_0, \tilde{T}_1, \dots, \tilde{T}_{N-1})^\intercal
\end{equation}
with respect to the temperature $T_\mathrm{env}$ of the environment. Hence, $T_i = \tilde{T}_i + T_\mathrm{env}$. The vector $\mathbf{B}$ solely contains the external heat transfer rates
\begin{equation}
    \mathbf{B} = (\dot{Q}_0, \dot{Q}_1, \dots, \dot{Q}_{N-1})^\intercal \mathrm{.}
    \label{eq:b_q}
\end{equation}
Matrix $\mathbf{A}(\mathbf{x})$ is assembled as
\begin{equation}
    \mathbf{A}(\mathbf{x})= \frac{1}{R_\mathrm{env}}\mathds{1} + \sum_{(i,j)\in \mathbf{E}} \frac{x_{ij}}{R_{ij}}\mathbf{U}_{ij}\mathrm{,}
    \label{eq:Amain}
\end{equation}
where $R_\text{env}$ is the thermal resistance connecting to the environment and $\mathbf{U}_{ij}$ contains only four nonzero values:
\begin{equation}
    \mathbf{U}_{ij} = 
    \begin{pNiceMatrix}[first-row,first-col]
        & &\color{blue}\underset{\downarrow}{i}& &\color{blue}\underset{\downarrow}{j}& \\
        & &\vdots& &\vdots& \\
        \color{blue} i \rightarrow &\cdots&1&\cdots&-1&\cdots\\
        & &\vdots& &\vdots& \\
        \color{blue} j \rightarrow &\cdots&-1&\cdots&1&\cdots\\
        & &\vdots& &\vdots& \\
    \end{pNiceMatrix}
    \mathrm{.}
    \label{eq:U_ij}
\end{equation}
A detailed derivation of the linear system is given in Appendix \ref{sec:ls}. Solving the linear system of Eq.\,\eqref{eq:ls} produces results consistent with those from steady-state simulations using engineering tools, such as Modelica-based software \cite{fritzson_openmodelica_2020}. We demonstrate that in Appendix \ref{sec:modelica}. 

\section{Quantum Simulation-Based Optimization}\label{sec:QuSO}
In conventional simulation-based optimization \cite{wang_simulation_2013, nguyen_review_2014}, the optimal system configuration is determined by iteratively simulating the system and adjusting its configuration based on the obtained simulation results. In the simple example shown in Fig.\,\ref{fig:problem}\,(b), we are interested in the best system configuration $\mathbf{x}$ to achieve the lowest possible temperature at the battery $T_0$, for example. Here, the computations are relatively inexpensive. \changemarker{From an engineering perspective, limiting the scope to only heat conduction represents a meaningful initial step before adding more complex effects such as convection and radiation. Nonetheless, realistic thermal management systems involve vast design spaces with dozens of components (pumps, valves, heat exchangers) and complex fluid dynamics governed by the non-linear Navier-Stokes equations \cite{ferziger_computational_2020}. Scaling the framework to such systems would require two main extensions: First, the system size $N$ simply increases. Second, while standard solvers often address non-linearities via iterative linearization, the most computationally intensive step remains solving a large, sparse linear system at each iteration. QuSO is positioned to accelerate exactly this linear solution step within such a hybrid quantum-classical iterative loop.}

We can potentially reduce the computational workload by employing the QuSO algorithm proposed in Ref.\,\cite{stein_exponential_2024}. Therefore, we allow a superposition of configurations and thus their simultaneous evaluations using the Quantum Approximate Optimization Algorithm (QAOA) \cite{farhi_quantum_2014, hadfield_quantum_2019}. For the simulation task, we have the choice of various quantum linear system solvers. Within this work, we use the Quantum Singular Value Transform (QSVT) \cite{gilyen_quantum_2019, martyn_grand_2021} to invert matrix $\mathbf{A}$ and consequently solve the linear system in Eq.\,\eqref{eq:ls} by computing $\mathbf{\tilde{T}} = \mathbf{A}^{-1}\mathbf{B}$. The potentially reduced runtime complexity of the QSVT algorithm results in another source of speedup for the simulation-based optimization procedure. To encode $\mathbf{A}$ on the quantum computer, we use block-encoding techniques such as the Linear Combination of Unitaries (LCU) method \cite{childs_hamiltonian_2012}. Once the temperatures are encoded in the amplitudes of a quantum state, we employ Quantum Amplitude Estimation (QAE) \cite{lomonaco_quantum_2002} to extract the specific temperature of interest. This step is followed by Quantum Phase Application (QPA) \cite{stein_exponential_2024}, which maps the temperature into the phase as a cost term, thereby mimicking the action of a cost Hamiltonian.

Figure \ref{fig:QuSO_overview} illustrates the overall structure of the algorithm and its key quantum subroutines. In the following, we briefly introduce each subroutine and its particular implementation to solve our QuSO problem of optimizing the simplified cooling system. 

\subsection{Quantum Approximate Optimization Algorithm}
\begin{figure}[h]
    \begin{quantikz}[wire types={b}, classical gap=0.07cm]
        \lstick{$\ket{+}^{\otimes m}_\texttt{c}$}&\gate{e^{-i\gamma H_\mathrm{C}}} \gategroup[1, steps=2, style={dashed,rounded corners, fill=cyan!20, inner xsep=2pt},background]{$\circlearrowleft$ $p$ layers} &\gate{e^{-i\beta H_\mathrm{M}}}&\meter[style={draw=magenta}]{\textcolor{magenta}{\langle H_\mathrm{C} \rangle}}
    \end{quantikz}
    \tikz[overlay,remember picture] {\draw[-{Stealth[length=2mm]}, thick, draw=magenta] (-0.52,-0.15) to[out=-90,in=0] (-1.17,-0.4) node[anchor=west, below, text=magenta]{update parameters};}
    \caption{QAOA circuit with variational parameters $\gamma$, $\beta$ and Hamiltonians $H_\mathrm{C}$ and $H_\mathrm{M}$. The blue section highlights a single QAOA layer which is repeated $p$ times with individual parameters. A final measurement of $\langle H_\mathrm{C} \rangle$ is used to classically optimize $\gamma$ and $\beta$. After several iterations of that procedure, the circuit prepares the desired ground state of $H_\mathrm{C}$.}
    \label{fig:qaoa}
\end{figure}
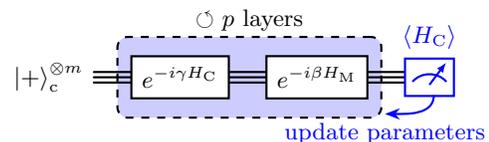

QAOA is designed to solve combinatorial optimization problems \cite{farhi_quantum_2014, hadfield_quantum_2019}. It is inspired by the Adiabatic Theorem \cite{born_beweis_1928}, which states that if a quantum system is initialized in the ground state of a certain initial Hamiltonian and evolves slowly enough under a time-dependent Hamiltonian, the system will remain in its instantaneous ground state, eventually reaching the ground state of the Hamiltonian of interest. Instead of continuous evolution, QAOA implements a discrete, gate-based approach. The algorithm begins by initializing all qubits to the $\ket{+}$ state, representing the ground state of the mixer Hamiltonian
\begin{equation}
    H_\mathrm{M} = \sum_{i=0}^{m-1} -X_i\mathrm{,}
\end{equation}
where $X_i$ is the Pauli-X operator applied to the $i$-th qubit. The optimization problem is encoded in the cost Hamiltonian $H_\mathrm{C}$. By alternately applying $p$ layers of the time evolutions $U_\mathrm{C}(\gamma) = e^{-i\gamma H_\mathrm{C}}$ and $U_\mathrm{M}(\beta) = e^{-i\beta H_\mathrm{M}}$, QAOA corresponds to a Suzuki-Trotter approximation \cite{suzuki_decomposition_1985, huyghebaert_product_1990} of the adiabatic evolution. For each layer, the parameters $\gamma$ and $\beta$ are classically updated to minimize the expectation value $\langle H_\mathrm{C} \rangle$, which acts as a cost function. The circuit structure of QAOA is briefly illustrated in Fig.\,\ref{fig:qaoa}.

\noindent To solve a QuSO problem, however, we need to perform a simulation to compute the cost function. Hence, we build a quantum algorithm to implement the cost layer as 
\begin{equation}
    U_\mathrm{C}(\gamma)\ket{\mathbf{x}}_\texttt{c} = e^{-i\gamma c(\mathbf{x})}\ket{\mathbf{x}}_\texttt{c}\mathrm{.}
    \label{eq:Uc}
\end{equation}
Here, $c(\mathbf{x})$ is the scalar cost value computed by the simulation for configuration $\mathbf{x}$. This can be interpreted as a time evolution under the diagonal cost Hamiltonian
\begin{equation}
    H_\mathrm{C} = \sum_{\mathbf{x}} c(\mathbf{x})\ketbra{\mathbf{x}}{\mathbf{x}}_\texttt{c}\mathrm{.}
    \label{eq:HC}
\end{equation}

\noindent Throughout the remainder of this paper, the QAOA register storing the configuration is referred to as the ${\texttt{c}\text{-register}}$.

\subsection{Matrix Block-Encoding}
To build the cost layer of Eq.\,\eqref{eq:Uc}, we need to compute the cost $c(\mathbf{x})$ on the quantum computer based on the solution of the linear system in Eq.\,\eqref{eq:ls}. Therefore, we encode the respective matrix $\mathbf{A}(\mathbf{x})$ in the top-left block of a unitary operator
\begin{equation}
    U_\mathbf{A} = 
    \begin{pmatrix}
        \mathbf{A}(\mathbf{x}) & *\\
        *&*
    \end{pmatrix}\mathrm{,}
\end{equation}
which acts on additional ancilla qubits. Notably, $\mathbf{A}(\mathbf{x})$ depends on the configuration $\mathbf{x}$, which is stored in the $\texttt{c}$-register as $\ket{\mathbf{x}}_\texttt{c}$. If several configurations exist in superposition, we also need to prepare our matrix in superposition. 

We achieve that block-encoding by the LCU-method \cite{childs_hamiltonian_2012}. As one can deduce from its name, $\mathbf{A}(\mathbf{x})$ is considered as a linear combination of unitary operators $U_k$:
\begin{equation}
    \mathbf{A}(\mathbf{x}) = \sum_k \lambda_k U_k\quad\text{with}\quad \lambda_k>0\mathrm{.}
    \label{eq:lcu}
\end{equation}
Therefore, an operation $V$ prepares an ancilla register $\texttt{l}$ to
\begin{equation}
    V\ket{0}_\texttt{l}\ket{\psi}_\texttt{d} = C_\text{LCU}\sum_k \sqrt{\lambda_k} \ket{k}_\texttt{l}\ket{\psi}_\texttt{d}\mathrm{,}
\end{equation}
where $C_\text{LCU}=(\sum_k\lambda_k)^{-1/2}$ ensures normalization. The state preparation step is followed by an operation $M_U$ which consists of a series of unitaries $U_k$ acting on the data register $\texttt{d}$ and controlled on register $\texttt{l}$. Consequently, each coefficient $\sqrt{\lambda_k}$ ends up in front of $U_k$:
\begin{equation}
    M_UV\ket{0}_\texttt{l}\ket{\psi}_\texttt{d} = C_\text{LCU}\sum_k \ket{k}_\texttt{l}\sqrt{\lambda_k}U_k\ket{\psi}_\texttt{d}\mathrm{.}
\end{equation}
Finally, we only have to unprepare the ancilla register to obtain
\begin{equation}
    \underbrace{V^\dagger M_UV}_{=U_\mathrm{LCU}{}}\ket{0}_\texttt{l}\ket{\psi}_\texttt{d} = C_\text{LCU}^2\ket{0}_\texttt{l}\underbrace{\sum_k \lambda_kU_k}_{=\mathbf{A}(\mathbf{x})}\ket{\psi}_\texttt{d} + \ket{\dots}_\texttt{ld}\mathrm{.}
    \label{eq:lcuop}
\end{equation}
If we only consider the $\texttt{d}$-register associated with $\ket{0}_\texttt{l}$, we indeed apply the matrix $\mathbf{A}(\mathbf{x})$ to the $\ket{\psi}_\texttt{d}$ state. All other terms in Eq.\,\eqref{eq:lcuop} are irrelevant and thus not further specified. If one is only interested in a state proportional to $\mathbf{A}(\mathbf{x})\ket{\psi}_\texttt{d}$, one could add a postselection measurement step, such that the ancilla register collapses to $\ket{0}_\texttt{l}$. In our case, we omit the measurement as we require a coherent operation. 

\changemarker{If we compare Eq.\,\eqref{eq:lcu} with the definition of $\mathbf{A}(\mathbf{x})$ in Eq.\,\eqref{eq:Amain}, we can identify the static coefficients $\lambda_k$ as either $1/(2 R_\mathrm{env})$ or $1/R_{ij}$, and the unitaries $U_k$ as $\mathds{1}$ or $[\mathbf{U}_{ij}/2]_\mathrm{block}$.} The latter is the block-encoding of the non-unitary matrix $\mathbf{U}_{ij}/2$ in Eq.\,\eqref{eq:U_ij}. \changemarker{Note that using static coefficients $\lambda_k$ corresponds to a fully connected graph where all $x_{ij}=1$. To correctly encode $\mathbf{A}(\mathbf{x})$ for an arbitrary configuration, we add controls on the $\texttt{c}$-register to the selection operator. This ensures that the term associated with $R_{ij}$ is only applied if the corresponding design variable $x_{ij}$ is set to $1$. More details follow lateron.}

While the implementation of $\mathds{1}$ is straightforward, we explain the construction of $[\mathbf{U}_{ij}/2]_\mathrm{block}$ in the following. Central to this construction is the LCU of
\begin{equation}
    \frac{1}{2}\mathds{1} - \frac{1}{2}X = \frac{1}{2}\begin{pmatrix}
        1 & -1 \\
        -1 & 1
    \end{pmatrix}\mathrm{,}\label{eq:x-1}
\end{equation}
which acts on a single qubit and requires one ancilla qubit initialized in an equal superposition state via a Hadamard gate ($V = H$). 

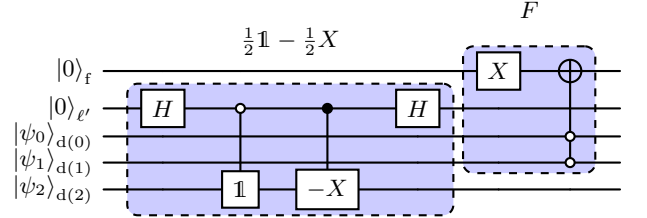
\begin{figure}[h]
    \begin{quantikz}[row sep=0.2cm]
        \lstick{$\ket{0}_\texttt{f}$} &  &  &  &  & \gate{X} \gategroup[4, steps=2, style={dashed,rounded corners, fill=cyan!20, inner xsep=2pt, inner ysep=-1pt}, background, label style={label={above:$F$}}]{} & \targ{} & \\[-0.2cm]
        \lstick{$\ket{0}_\texttt{l'}$} & \gate{H} \gategroup[4, steps=4, style={dashed,rounded corners, fill=cyan!20, inner xsep=2pt, inner ysep=-1pt}, background, label style={label={above:$\frac{1}{2}\mathds{1} - \frac{1}{2}X$}}]{} & \octrl{3} & \ctrl{3} & \gate{H} & & & \\[-0.15cm]
        \lstick{$\ket{\psi_0}_\texttt{d}$} & & & & & & \octrl{-2 }& \\
        \lstick{$\ket{\psi_1}_\texttt{d}$} & & & & & & \octrl{-1} & \\[-0.18cm]
        \lstick{$\ket{\psi_2}_\texttt{d}$} & & \gate{\mathds{1}} & \gate{-X} & & & & 
    \end{quantikz}
    \caption{Quantum circuit to embed the LCU of Eq.\,\eqref{eq:x-1} in the top-left block of an $8\times8$ zero-padded matrix. This corresponds to $[\mathbf{U}_{01}/2]_\mathrm{block}$, and d(2) represents the least significant qubit.}
    \label{fig:lcu_upperleftblock}
\end{figure}

We first embed this $2\times2$ matrix into the top-left block of an $N\times N$ zero-padded matrix, effectively constructing $[\mathbf{U}_{01}/2]_\mathrm{block}$. To achieve this, we add $\lceil\log_2N\rceil-1$ qubits to the $\texttt{d}$-register and introduce a flag qubit $\texttt{f}$ and an operation $F$, which flips $\texttt{f}$ iff all but the least significant data qubits are $\ket{0}$.
By treating $\texttt{f}$ as an ancilla qubit of the block-encoding, residual values are shifted out of the top-left block. The resulting circuit for $N=8$ is shown in Fig.\,\ref{fig:lcu_upperleftblock}.

To generalize this to arbitrary indices $i$ and $j$, we permute the basis states such that 
\begin{align}
\begin{split}
    \ket{i}_\texttt{d}=\ket{i_0i_1\dots i_{n-1}}\,\mapsto&\ket{0}=\ket{00\dots0}_\texttt{d}\\
    \ket{j}_\texttt{d}=\ket{j_0j_1\dots j_{n-1}}\mapsto&\ket{1}=\ket{00\dots1}_\texttt{d}\mathrm{,}
\end{split}
\label{eq:perm}
\end{align}
where $n=\lceil\log_2N\rceil$.
This is achieved by first flipping the qubits where the binary representation of $i$ has a 1, resulting in $\ket{i} \mapsto \ket{0}$ and $\ket{j} \mapsto \ket{j\oplus i}$. We then use CNOT gates controlled on an arbitrarily chosen qubit corresponding to a 1 in the binary representation of $j\oplus i$ applied on the remaining 1's. A final SWAP operation moves the qubit with the remaining 1 to the least significant position, meaning $\ket{j}\mapsto\ket{1}$. We term the entire permutation in Eq.\,\eqref{eq:perm} $P_{ij}$.

The block-encoding of $\mathbf{U}_{ij}/2$ is thus given by
\begin{equation}
    \left[\frac{1}{2}\mathbf{U}_{ij}\right]_\mathrm{block} = P_{ij}^\dagger F\left[\frac{1}{2}\mathds{1} - \frac{1}{2}X\right]_\mathrm{LCU}P_{ij}\mathrm{,}
    \label{eq:PUPdagger}
\end{equation}
where the subscript LCU indicates that the expression is implemented with the LCU method.

The worst-case scenario for the implementation of the $P_{ij}$ operation in terms of gate depth occurs when $i = (11\ldots1)_2$ and $j = (00\ldots0)_2$. This case requires $\lceil \log_2 N \rceil$ $X$ gates, $\lceil \log_2 N \rceil - 1$ CNOT gates, and one SWAP gate. Consequently, the algorithmic complexity of block-encoding $\mathbf{U}_{ij} / 2$ in Eq.~\eqref{eq:PUPdagger} is $\mathcal{O}(\log N)$.
The entire LCU-based construction of $\mathbf{A}(\mathbf{x})$ is given by
\begin{equation}
    U_\mathbf{A} = \left[\frac{C_\text{LCU}}{2R_\mathrm{env}}\mathds{1} + C_\text{LCU} \sum_{(i, j) \in \mathbf{E}} \frac{x_{ij}}{R_{ij}} \left[\frac{1}{2} \mathbf{U}_{ij} \right]_\mathrm{block}\right]_\mathrm{LCU},
    \label{eq:UA}
\end{equation}
which comprises $m + 1$ terms. Preparing these $m + 1$ amplitudes typically requires $\mathcal{O}(m)$ operations~\cite{barenco_elementary_1995, zhang_quantum_2022, grover_creating_2002, ozols_quantum_2013}. Additionally, there are $m$ controlled unitaries, resulting in an overall gate complexity for $U_\mathbf{A}$ of
\begin{equation}
    \mathcal{G}_\mathrm{LCU} = \mathcal{O}(m \log N).
    \label{eq:G_LCU_1}
\end{equation}

\begin{figure}[h]
    \begin{quantikz}[row sep=0.1cm]
        \lstick{$\ket{\mathbf{x}}_\texttt{c}$} & & & \phase{x_{ij}} \wire[d][2]{q} \gategroup[4, steps=1, style={dashed,rounded corners, fill=cyan!20},background]{$\forall(i, j)\in\mathbf{E}$} & & \\[-0.1cm]
        \lstick{$\ket{0}_\texttt{l}$} & \gate{V} & \octrl{1} & \uctrl{} \wire[d][1]{q} & \gate{V^\dagger} &\\[-0.22cm]
        \lstick{$\ket{0}_\texttt{fl'}$} & & \gate[2]{\mathds{1}} & \gate[2, disable auto height]{\left[\frac{1}{2}\mathbf{U}_{ij}\right]_\mathrm{block}} & & \\[-0.2cm]
        \lstick{$\ket{0}_\texttt{d}$} & & & & &
    \end{quantikz}
    \caption{Quantum circuit for the block-encoding $U_\mathbf{A}$ of $\mathbf{A}(\mathbf{x})$ (cf. Eq.\,\eqref{eq:UA}). The block-encoding of $\mathbf{U}_{ij}/2$ is not only controlled on the $\texttt{l}$-register (indicated by the half-filled control node) but also on the respective qubit $\ket{x_{ij}}$ of the ${\texttt{c}\text{-register}}$. The operation within the blue shaded area is repeated for all connections $(i, j)\in\mathbf{E}$.}
    \label{fig:UA}
\end{figure}
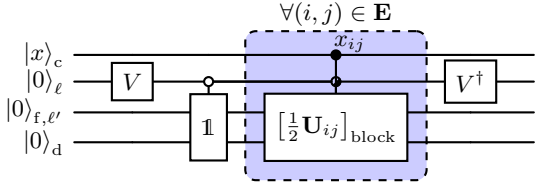

Since some coefficients in Eq.\,\eqref{eq:UA} explicitly depend on the binary variables $x_{ij}$ and therefore on the configuration $\mathbf{x}$, we add controls on the ${\texttt{c}\text{-register}}$. Consequently, if multiple configurations exist in superposition, the resulting circuit achieves a coherent encoding of $\mathbf{A}(\mathbf{x})$ in superposition. The corresponding quantum circuit is shown in Fig.\,\ref{fig:UA}.

For a graph with $N$ nodes, the number of edges $m$ in a connected graph is bounded by ${N-1 \leq m \leq N(N-1)/2}$. In practice, we anticipate sparse graphs for realistic systems involving dozens of components. Given that the maximum number of pipes that can be connected to a node is physically constrained, we assume a finite maximum degree of the graph ${d_\mathrm{max} = \mathcal{O}(1)}$. This leads to an upper bound of 
\begin{equation}
    m = \frac{1}{2}\sum_{i=0}^{N-1} d_i \leq \frac{1}{2}N d_\mathrm{max},
\end{equation}
where $d_i$ represents the degree of node $i$. 
Using this sparsity condition, we express the gate complexity in Eq.~\eqref{eq:G_LCU_1} in terms of $N$ as
\begin{equation}
    \mathcal{G}_\mathrm{LCU} = \mathcal{O}(N \log N).
    \label{eq:G_LCU_2}
\end{equation}

The whole matrix construction requires $\lceil \log_2(m+1) \rceil$ ancilla qubits in the $\texttt{l}$-register for the outer LCU layer, one $\texttt{l}'$ qubit for the inner LCU layer, one $\texttt{f}$ qubit, and $\lceil \log_2 N \rceil$ $\texttt{d}$ qubits.

Due to the normalization condition and the LCU procedure, the resulting block-encoded matrix corresponds to $C_\text{LCU}^2\mathbf{A}(\mathbf{x})/2$. 

\subsection{Quantum Singular Value Transform}\label{sec:qsvt}
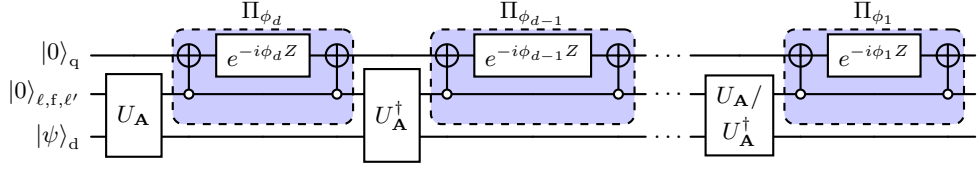
\begin{figure*}[hbt!]
    \begin{quantikz}[row sep=0.1cm, column sep=0.2cm]
        \lstick{$\ket{0}_\texttt{q}$}              & & \targ{} \gategroup[2, steps=3, style={dashed,rounded corners, fill=cyan!20, inner xsep=2pt, inner xsep=-0.05cm, inner ysep=-0.05cm}, background]{$\Pi_{\phi_d}$} & \gate{e^{-i\phi_dZ}} & \targ{} & & \targ{} \gategroup[2, steps=3, style={dashed,rounded corners, fill=cyan!20, inner xsep=2pt, inner xsep=-0.05cm, inner ysep=-0.05cm}, background]{$\Pi_{\phi_{d-1}}$} & \gate{e^{-i\phi_{d-1}Z}} & \targ{} & \ \ldots\  & & \targ{} \gategroup[2, steps=3, style={dashed,rounded corners, fill=cyan!20, inner xsep=2pt, inner xsep=-0.05cm, inner ysep=-0.05cm}, background]{$\Pi_{\phi_1}$} & \gate{e^{-i\phi_1Z}} & \targ{} &\\[-0.2cm]
        \lstick{$\ket{0}_\texttt{lfl'}$}   & \gate[3]{U_\mathbf{A}} & \octrl{-1} & & \octrl{-1} & \gate[3]{U_\mathbf{A}^\dagger} & \octrl{-1} & & \octrl{-1} & \ \ldots\  & \gate[3, disable auto height]{\shortstack{$U_\mathbf{A}$\text{/}\\ $U_\mathbf{A}^\dagger$}} & \octrl{-1} & & \octrl{-1} &\\
        \lstick{$\ket{\psi}_\texttt{d}$}           & & & & & & & & & \ \ldots\  & & & & & \\
        \lstick{$\ket{\mathbf{x}}_\texttt{c}$}           & & & & & & & & & \ \ldots\  & & & & & 
    \end{quantikz}
    \caption{Quantum circuit representation of the Quantum Singular Value Transformation (QSVT) operator. The unitary $U_\mathbf{A}$ denotes the block-encoding of the matrix $\mathbf{A}(\mathbf{x})$, and the blue-shaded areas represent projector-controlled phase-shift operators. For even values of $d$, the rightmost block-encoding operator $U_\mathbf{A}$ should be replaced with its Hermitian adjoint $U_\mathbf{A}^\dagger$. To improve readability, we have omitted the controls on the ${\texttt{c}\text{-register}}$.}
    \label{fig:qsvt}
\end{figure*}
The Quantum Singular Value Transform (QSVT) is a quantum algorithm that manipulates the singular values of a matrix encoded within a quantum circuit \cite{gilyen_quantum_2019, martyn_grand_2021}. By doing so, it is capable of solving similar problems such as Quantum Phase Estimation (QPE) \cite{nielsen_quantum_2012}, Grover's search \cite{grover_fast_1996}, Hamiltonian simulation \cite{low_optimal_2017}, and matrix inversion \cite{martyn_grand_2021}. The algorithm applies a $d$-degree polynomial transformation on the singular values $P^\mathrm{sv}(\mathbf{A})$ of matrix $\mathbf{A} = \mathbf{W}_\Sigma\mathbf{\Sigma}\mathbf{V}_\Sigma^\dagger$, where $\mathbf{\Sigma}$ contains the singular values in its diagonal. For our square matrix, the QSVT operator is defined as
\begin{align}
\begin{split}
    U_{\boldsymbol{\phi}} &= 
    \begin{cases}
        \Pi_{\phi_1}U_\mathbf{A}\prod\limits_{k=1}^{(d-1)/2}\Pi_{\phi_{2k}}U_\mathbf{A}^\dagger\Pi_{\phi_{2k+1}}U_\mathbf{A}\text{,}&\text{for odd }d \\
        \prod\limits_{k=1}^{d/2}\Pi_{\phi_{2k-1}}U_\mathbf{A}^\dagger\Pi_{\phi_{2k}}U_\mathbf{A}\text{,}&\text{for even }d
    \end{cases}\\
    &=
    \begin{pmatrix}
        \begin{array}{
          w{c}{\widthof{$\mathbf{W}_\Sigma P(\mathbf{\Sigma})\mathbf{V}_\Sigma^\dagger$}}
          w{c}{\widthof{$\mathbf{W}_\Sigma P(\mathbf{\Sigma})\mathbf{V}_\Sigma^\dagger$}}
          w{c}{\widthof{$\mathbf{W}_\Sigma P(\mathbf{\Sigma})\mathbf{V}_\Sigma^\dagger$}}
          w{c}{\widthof{$\mathbf{W}_\Sigma P(\mathbf{\Sigma})\mathbf{V}_\Sigma^\dagger$}}
        }
        P^\mathrm{sv}(\mathbf{A}) & *\\
        * & *
    \end{array}
    \end{pmatrix}\mathrm{.}
\end{split}
\end{align}
Here, $U_\mathbf{A}$ is the block-encoding of $\mathbf{A}(\mathbf{x})$ as defined in Eq.\,\eqref{eq:UA}, and $\Pi_{\phi_i}$ is a projector-controlled phase-shift operation given by $\Pi_{\phi_i}=e^{i\phi_i(2\Pi-\mathds{1})}$, where ${\Pi=\ketbra{0}{0}_\texttt{lfl'}}$ is the projector associated with our block-encoding \footnote{The QSVT operator in some literature includes an additional projector-controlled phase-shift operation. This introduces an extra phase angle $\phi_0$ or $\phi_{d+1}$, which is required to transition between different conventions of block-encoding.}. The corresponding quantum circuit in Fig.~\ref{fig:qsvt} illustrates how $\Pi_{\phi_i}$ can be constructed using an ancilla qubit $\texttt{q}$, two CNOT gates, and a $Z$ phase gate.

Within the QuSO algorithm, we employ the QSVT routine to invert $\mathbf{A}(\mathbf{x})$, thereby solving the linear system in Eq.\,\eqref{eq:ls}. Therefore, we choose the phase angles $\boldsymbol{\phi}=(\phi_1, \phi_2, \dots, \phi_d)$ such that the resulting polynomial approximates the reciprocal function ${P_\mathrm{inv}(x_\mathrm{p})\approx C_\mathrm{p}/x_\mathrm{p}}$, where $C_\mathrm{p}$ ensures that the function values lie within $[-1, 1]$ \cite{dong_efficient_2021, martyn_grand_2021, gilyen_quantum_2019}. In combination with the block-encoding of $\mathbf{A}^\dagger$, the resulting operator approximately block-encodes $\mathbf{A}^{-1}$ up to a constant factor:
\begin{align}
\begin{split}
    P_\mathrm{inv}^\mathrm{sv}\left(\mathbf{A}^\dagger\right) &= \mathbf{V}_\Sigma P_\mathrm{inv}(\mathbf{\Sigma})\mathbf{W}_\Sigma^\dagger \\
    &\approx C_\mathrm{p} \mathbf{V}_\Sigma\mathbf{\Sigma}^{-1}\mathbf{W}_\Sigma^\dagger = C_\mathrm{p}\mathbf{A}^{-1}.
\end{split}
\end{align}
To implement this inversion, we use the \textit{matrix-inversion polynomial} $P^\mathrm{MI}_{\epsilon, \mu}(x_p)$ introduced in Ref.\,\cite{martyn_grand_2021}. This polynomial is an $\epsilon\mu/2$-close approximation to $(\mu/2)/x_\mathrm{p}$ for $x_\mathrm{p}\in[-1, 1]\setminus[-\mu, \mu]$. Thus, $C_\mathrm{p}=\mu/2$, where $\mu$ is the threshold value corresponding to the smallest invertible singular value and thus satisfies $\mu \leq \sigma_\text{min} C_\text{LCU}^2/2\leq1/\kappa$ \footnote{In Ref.\,\cite{martyn_grand_2021}, the authors set $\mu = 1/\kappa$ under the assumption that the largest singular value $\sigma_{\max}=1$. Since this is not true in general, we introduce $\mu$ to avoid confusion with the condition number $\kappa$.}, and $\epsilon$ controls the accuracy. Here, $\sigma_\mathrm{min}$ denotes the smallest singular value of $\mathbf{A}(\mathbf{x})$, and $\kappa$ is its condition number. The polynomial has odd parity and its degree scales as $\mathcal{O}(\kappa\log(\kappa/\epsilon))$, resulting in a similar number of calls to the block-encoding $U_\mathbf{A}$. Since we know the exact structure of $\mathbf{A}(\mathbf{x})$ for all configurations $x$, we can derive an asymptotic scaling of $\kappa = \mathcal{O}(R_\mathrm{env}/R_\mathrm{min})$, where $R_\mathrm{min}$ is the minimum thermal resistance of the system. A detailed derivation can be found in Appendix \ref{sec:matrixprops}. Consequently, the QSVT circuit has a gate complexity of 
\begin{equation}
    \mathcal{G}_\mathrm{QSVT} = \mathcal{O}\left(\frac{R_\mathrm{env}}{R_\mathrm{min}}\log\left(\frac{R_\mathrm{env}}{\epsilon R_\mathrm{min}}\right)\right) \times \mathcal{G}_\mathrm{LCU}.
    \label{eq:G_QSVT}
\end{equation}
In order to obtain the temperature vector ${\tilde{\mathbf{T}}(\mathbf{x}) = \mathbf{A}^{-1}(\mathbf{x})\mathbf{B}}$, we need to prepare $\mathbf{B}$ in the data register prior to applying the QSVT operator:
\begin{equation}
    V_\mathrm{B}\ket{0}_\texttt{d} = C_\mathrm{B} \sum_{k=0}^{N-1}\dot{Q}_k \ket{k}_\texttt{d}.
\end{equation}
Here, $C_\mathrm{B} = (\sum_k\dot{Q}_k^2)^{-1/2}$ ensures normalization. In realistic systems, the number of external heat and cooling sources $N_\mathrm{Q}$ is much smaller than $N$, as not every component produces heat or actively cools the system. As a result, the vector $\mathbf{B}$ is sparse, containing only $N_\mathrm{Q}$ nonzero values. This sparsity reduces the gate complexity for the state preparation operation $V_\mathrm{B}$ from $\mathcal{O}(N)$ to $\mathcal{O}(N_\mathrm{Q} \log N)$ \cite{dalzell_quantum_2025, ramacciotti_simple_2024}. Since $\mathcal{G}_\mathrm{LCU}$ in Eq.\,\eqref{eq:G_LCU_2} dominates, the overall gate complexity of the linear system solving routine $L = U_{\boldsymbol{\phi}}V_\mathrm{B}$ remains equivalent to $\mathcal{G}_\mathrm{QSVT}$ in Eq.\,\eqref{eq:G_QSVT}.

After applying $L$, we obtain a state proportional to $\tilde{\mathbf{T}}(\mathbf{x})$:
\begin{align}
\begin{split}
    L&\ket{\mathbf{x}}_\texttt{c}\ket{0}_\texttt{qlfl'}\ket{0}_\texttt{d}\\
    &= U_{\boldsymbol{\phi}}\left(\ket{\mathbf{x}}_\texttt{c}\ket{0}_\texttt{qlfl'}V_\mathrm{B}\ket{0}_\texttt{d}\right)\\
    & = \frac{2}{C_\text{LCU}^2}C_\mathrm{p}C_\mathrm{B} \ket{\mathbf{x}}_\texttt{c}\ket{0}_\texttt{qlfl'} \underbrace{\sum_{k=0}^{N-1} \tilde{T}_k \ket{k}_\texttt{d}}_{=\tilde{\mathbf{T}}(\mathbf{x})} + \ket{\dots}_\texttt{cqlfl'd}.
\end{split}
\label{eq:L}
\end{align}

\subsection{Quantum Amplitude Estimation}
\begin{figure}[h!]
    \begin{quantikz}[row sep = 0.1cm, column sep = 0.3 cm]
        \lstick{$\ket{0}_\texttt{p}$} & \gate{H} & \ctrl{3} & & & \gate[3]{\mathrm{QFT}^\dagger} & \rstick{$\ket{\theta_0}$}\\
        \lstick{$\ket{0}_\texttt{p}$} & \gate{H} & & \ctrl{2} & & & \rstick{$\ket{\theta_1}$}\\
        \lstick{$\ket{0}_\texttt{p}$} & \gate{H} & & & \ctrl{1} & & \rstick{$\ket{\theta_2}$}\\
        \lstick{$\ket{0}_\texttt{qlfl'dc}$} & \gate{L} & \gate{G} & \gate{G^2} & \gate{G^4} & & 
    \end{quantikz}
    \caption{Quantum circuit for quantum amplitude estimation without measurements. To improve readability, we have omitted the controls on the ${\texttt{c}\text{-register}}$.}
    \label{fig:qae}
\end{figure}
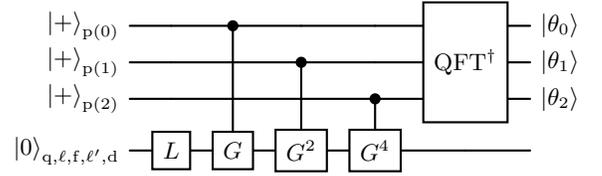

As the name Quantum Amplitude Estimation (QAE) \cite{lomonaco_quantum_2002} suggests, it is an algorithm to extract the absolute value of a quantum amplitude $a$ of some state 
\begin{equation}
    \ket{\phi} = a\ket{\alpha} + \sqrt{1-a^2}\ket{\alpha^\perp}.
\end{equation}
Therefore, it uses Quantum Phase Estimation (QPE) \cite{nielsen_quantum_2012} (cf. Fig.\,\ref{fig:qae}) with the Grover operator
\begin{equation}
    G = -LS_0L^\dagger S_\alpha,
    \label{eq:grover}
\end{equation}
where $L$ prepares the solution of the linear system, ${S_0 = \mathds{1}-2\ketbra{0}{0}}$ performs a reflection about the $\ket{0}$ state, and ${S_\alpha = \mathds{1}-2\ketbra{\alpha}{\alpha}}$ reflects about the $\ket{\alpha}$ state.

\noindent In the subspace spanned by $\ket{\alpha}$ and $\ket{\alpha^\perp}$, the Grover operator has two eigenstates
\begin{equation}
    \ket{\psi_\pm} = \frac{1}{\sqrt{2}}(\ket{\alpha}\pm i\ket{\alpha^\perp})
    \label{eq:psi}
\end{equation}
with the corresponding eigenvalues
\begin{equation}
    \lambda_\pm = e^{\pm i2\pi\theta},
    \label{eq:eigvals}
\end{equation}
where the phase $\theta$ is related to the amplitude $a$ by
\begin{equation}
    \abs{a} = \sin(\pi\theta).
\end{equation}
At the end of the QAE circuit, the phase register p, consisting of $k$ qubits, encodes the binary fraction of ${\theta = (0.\theta_0\theta_1\dots\theta_{k-1})_2}$. The precision of $\theta$ is determined by the error $\delta=2^{-k}$, resulting in a gate complexity for QAE of
\begin{equation}
    \mathcal{G}_\mathrm{QAE} = \mathcal{O}\left(\frac{1}{\delta}\right)\times\mathcal{G}_\mathrm{QSVT}.
\end{equation}

In our QuSO algorithm, we compute the temperature vector $\tilde{\mathbf{T}}(\mathbf{x})$ in the subspace of data register $\texttt{d}$, where all ancilla qubits are in the $\ket{0}$ state. We employ QAE to extract the temperature of interest $\tilde{T}_{\alpha}$, which corresponds to the cost function we aim to minimize. Within this work, we assume that $\tilde{T}_{\alpha} > 0$ to avoid the need for representing negative numbers with binary fractions. If this condition is not met, two's complement can be used instead \cite{stein_exponential_2024}. At the end, the phase is given by
\begin{equation}
    \theta(\mathbf{x}) = \arcsin(2C_\mathrm{p}C_\mathrm{B}\tilde{T}_{\alpha}(\mathbf{x})/C_\text{LCU}^2)/\pi.
    \label{eq:theta1}
\end{equation}
\begin{figure*}[hbt!]
    \centering
    \includegraphics{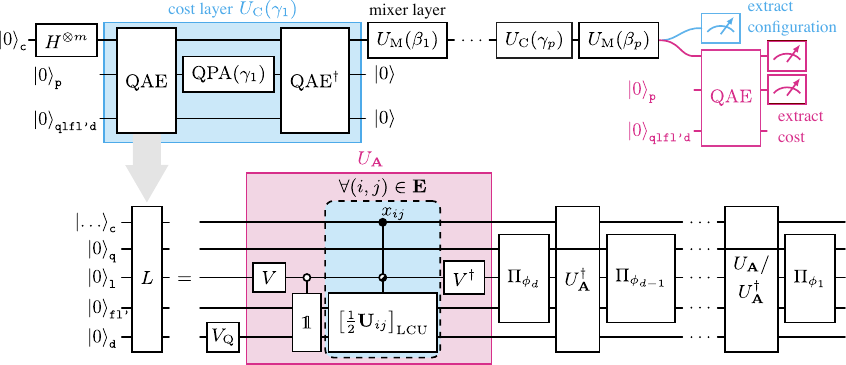}
    \caption{Overview of the QuSO circuit. The upper part shows the QAOA circuit, starting with Hadamard gates and alternating cost and mixer layers. Each cost layer includes QAE to extract the cost into the phase register p, QPA($\gamma$) to apply it as a phase, and QAE$^\dagger$ to disentangle ancilla qubits. The ${\texttt{c}\text{-register}}$ is measured to extract the optimal bitstring, and an additional QAE run retrieves its associated cost. The lower part illustrates the quantum linear system solver $L$, comprising state preparation $V_\mathrm{B}$ for encoding $\mathbf{B}$ and QSVT for inverting $\mathbf{A}$. Nested LCUs in the block-encoding $U_\mathbf{A}$ have controls on $\ket{x_{ij}}$ states in the ${\texttt{c}\text{-register}}$, ensuring inclusion only for valid terms. Half-filled control nodes represent one or more multi-controlled operations.}
    \label{fig:QuSO_overview}
\end{figure*}
\subsection{Quantum Phase Application}
The phase register $\texttt{p}$ encodes the phase $\theta(\mathbf{x})$ from Eq.\,\eqref{eq:theta1}. The Quantum Phase Application (QPA) routine performs
\begin{equation}
    \mathrm{QPA}(\gamma)\ket{\theta}_\texttt{p} = e^{-i\gamma f(\theta)}\ket{\theta}_\texttt{p},
\end{equation}
where $f(\theta)=\sin(\pi\theta)$ such that we obtain the cost layer of Eq.\,\eqref{eq:Uc} with $c(\mathbf{x})=\tilde{T}_{\alpha}(\mathbf{x})\cdot 2C_\mathrm{p}C_\mathrm{B}/C_\text{LCU}^2$. For a phase register of size $k$, $f(\theta)$ can be implemented exactly by constructing a polynomial of degree up to $2^k$ that matches $f(\theta)$ at all $2^k$ discrete inputs \cite{collins_fundamental_2003}. The polynomial is realized using Pauli-$Z$ operators
\begin{align}
\begin{split}
    \mathrm{QPA}(\gamma)\ket{\theta}_\texttt{p} &= e^{-i\gamma \sum_{S \subseteq \{0, 1, \dots, k-1\}} a_S \bigotimes_{i \in S} Z_i}\\
    &= \prod_{S \subseteq \{0, 1, \dots, k-1\}} e^{-i\gamma a_S \bigotimes_{i \in S} Z_i},\label{eq:MZZ}
\end{split}
\end{align}
where $a_S$ are the $2^k$ polynomial coefficients. The optimal implementation of $\mathrm{QPA}(\gamma)$ requires
\begin{equation}
    \mathcal{G}_\mathrm{QPA} = \mathcal{O}\left(2^k\right) = \mathcal{O}\left(\frac{1}{\delta}\right)
    \label{eq:G_QPA}
\end{equation}
gates \cite{welch_efficient_2014}.

Since the sine function can be well-approximated using low-degree polynomials, the number of gates in Eq.\,\eqref{eq:MZZ} can be significantly reduced. However, as $\mathcal{G}_\mathrm{QAE}$ is typically the dominant cost, implementing an exact QPA circuit becomes negligible in the overall complexity.

\subsection{QuSO Algorithm}
In the previous sections, we introduced the vanilla QAOA algorithm for combinatorial optimization problems, outlined the encoding of matrix $\mathbf{A}(\mathbf{x})$ using nested LCU block-encodings, and described the use of QSVT to invert $\mathbf{A}(\mathbf{x})$ and solve the linear system in Eq.\,\eqref{eq:ls}. We further detailed the extraction of a single amplitude $\tilde{T}_{\alpha}$ from the solution vector $\tilde{\mathbf{T}}(\mathbf{x})$ via QAE, as well as the application of this amplitude as a global phase using QPA to implement the cost layer of the QAOA algorithm $U_\mathrm{C}(\gamma)$ from Eq.\,\eqref{eq:Uc}. The QuSO algorithm integrates all these quantum subroutines into a complete QAOA-based framework to find the optimal configuration of the system \cite{stein_exponential_2024}. The overall QuSO circuit is depicted in Fig.\,\ref{fig:QuSO_overview}.

The cost layer $U_\mathrm{C}(\gamma)$ is given by 
\begin{equation}
    U_\mathrm{C}(\gamma) = \mathrm{QAE}^\dagger \mathrm{QPA}(\gamma)\mathrm{QAE}.
    \label{eq:costlayer}
\end{equation}
where $\mathrm{QAE}$ computes the phase
\begin{equation}
    \theta(\mathbf{x}) = \pm\tfrac{1}{\pi}\arcsin(c(\mathbf{x})).
    \label{eq:theta}
\end{equation} 

Ideally, the $k$ qubits in the $\texttt{p}$-register are sufficient to represent $\theta(\mathbf{x})$ exactly, and the qubit registers $\texttt{pqlfl'd}$ are uncomputed to the $\ket{0}$ state after each cost layer. In general, however, $k$ might be insufficient, so the phases $\theta(\mathbf{x})$ are computed with an error of $\delta=2^{-k}$, and the ancilla qubits are not perfectly reset. The resulting error scales as $\mathcal{O}(\delta)$ and is thus negligible for small $\delta$. Further details are provided in Appendix \ref{sec:qae_precision}. 

After applying the $p$ cost and mixer layers, the optimal configuration can be extracted by sampling from the ${\texttt{c}\text{-register}}$, identifying the bitstring with the highest probability. This bitstring corresponds to the current best configuration of the system. For classical optimization, the cost of the identified configuration must also be evaluated, which can be achieved with an additional $\mathrm{QAE}$ execution and a subsequent measurement of the $\texttt{p}$-register. An additional curve fit could further increase the accuracy of the extracted cost \cite{lim_curve-fitted_2024}.

\changemarker{
By aggregating the gate complexities of all quantum subroutines, we can derive the QuSO gate complexity for a single circuit execution as
\begin{equation}
    \mathcal{G}_\text{QuSO} = \mathcal{O}\left(p\,N\log N \frac{R_\mathrm{env}}{\delta R_\mathrm{min}}\log\left(\frac{R_\mathrm{env}}{\epsilon R_\mathrm{min}}\right)\right),
    \label{eq:gate_QuSO}
\end{equation}
where we included the number of QAOA layers $p$. To infer the total algorithmic runtime complexity $\mathcal{R}_\text{QuSO}$, we must account for the number of circuit executions required by the classical optimizer. For QAOA, the objective function is the expectation value
\begin{equation}
    \langle H_\mathrm{C} \rangle = \sum_\mathbf{x} P(\mathbf{x})\,c(\mathbf{x)},
\end{equation}
where $P(\mathbf{x}) = |\braket{\mathbf{x}}{\psi(\boldsymbol{\gamma}, \boldsymbol{\beta})}_\texttt{c}|^2$ is the probability of measuring configuration $\mathbf{x}$. Since extracting the full probability distribution is inefficient, we estimate the expectation value via Monte Carlo sampling. We sample $n_\text{MC} = \mathcal{O}(\epsilon_\text{MC}^{-2})$ configurations from the $\texttt{c}$-register to achieve a statistical error of $\epsilon_\text{MC}$. For each sampled configuration $\mathbf{x}$, we must compute the cost $c(\mathbf{x})$ by running the QAE routine. Crucially, since the success probability of QAE is independent of the system size, reliable cost extraction requires only a constant number of shots per sample, leading to a negligible overhead.}

\changemarker{
The total runtime complexity depends on the optimization strategy. If we employ a gradient-free optimizer (e.g., COBYLA \cite{powell_direct_1994}), which typically requires one function evaluation per iteration, the runtime complexity is
\begin{equation}
    \mathcal{R}_\text{QuSO}^\text{nograd}=\mathcal{O}\left(\epsilon_\text{MC}^{-2}\,n_\text{iter}\,p\,N\log N \frac{R_\mathrm{env}}{\delta R_\mathrm{min}}\log\left(\frac{R_\mathrm{env}}{\epsilon R_\mathrm{min}}\right)\right).
    \label{eq:runtime_nograd}
\end{equation}
Note that we have added the number of iterations $n_\text{iter}$ within the classical optimizer.
Alternatively, a gradient-based optimizer using the parameter-shift rule requires calculating partial derivatives with respect to all $2p$ variational parameters. This requires $O(p)$ function evaluations per iteration, resulting in quadratic scaling with respect to depth $p$:
\begin{equation}
    \mathcal{R}_\text{QuSO}^\text{grad}=\mathcal{O}\left(\epsilon_\text{MC}^{-2}\,n_\text{iter}\,p^2\,N\log N \frac{R_\mathrm{env}}{\delta R_\mathrm{min}}\log\left(\frac{R_\mathrm{env}}{\epsilon R_\mathrm{min}}\right)\right).
    \label{eq:runtime_grad}
\end{equation}}
We do not further analyze the dependency of $p$ and $n_\text{iter}$ on system size $N$, since the choice of hyperparameters and optimization strategies has a significant impact. Table\,\ref{tab:complexities} provides an overview of the gate complexity of each algorithmic subroutine.

\begin{table}[H]
    \centering
    \renewcommand{\arraystretch}{1.5}
    \begin{adjustbox}{max width=\columnwidth}
    \begin{tabular}{c c c c}
        \toprule
        Algorithm & Gate Complexity $\mathcal{G}$ & Space Complexity $\mathcal{S}$ \\
        \midrule
        LCU & $\mathcal{O}(N\log N)$ & $\mathcal{O}(\log N)$ \\
        QSVT & $\mathcal{O}\left(\frac{R_\mathrm{env}}{R_\mathrm{min}}\log\left(\frac{R_\mathrm{env}}{\epsilon R_\mathrm{min}}\right)\right) \times \mathcal{G}_\mathrm{LCU}$ & $\mathcal{S}_\mathrm{LCU} + \mathcal{O}(1)$ \\
        QAE & $\mathcal{O}\left(\frac{1}{\delta}\right)\times \mathcal{G}_\mathrm{QSVT}$ & $\mathcal{S}_\mathrm{QSVT} + \mathcal{O}\left(\log\frac{1}{\delta}\right)$ \\
        QPA & $\mathcal{O}\left(\frac{1}{\delta}\right)$ & $\mathcal{O}\left(\log\frac{1}{\delta}\right)$ \\
        \midrule
        QuSO &$\mathcal{O}\left(p\,N\log N \frac{R_\mathrm{env}}{\delta R_\mathrm{min}}\log\left(\frac{R_\mathrm{env}}{\epsilon R_\mathrm{min}}\right)\right)$ & $\mathcal{O}\left(\max\left\{N, \log\frac{1}{\delta}\right\}\right)$ \\
        \bottomrule
    \end{tabular}
    \end{adjustbox}
    \caption{Gate and space complexities, as well as total qubit counts for QuSO and its subroutines. \changemarker{Resulting runtime complexities are given in Eqs.\,\eqref{eq:runtime_nograd} and \eqref{eq:runtime_grad}.}}
    \label{tab:complexities}
\end{table}

\changemarker{The QAE approximation error $\delta$ and the QSVT approximation error $\epsilon$ equally impact the precision of the computed cost value $c(\mathbf{x})$. This is because $\delta$ directly sets the resolution of $c(\mathbf{x})$ and $\epsilon$ sets the precision of the solution vector $\mathbf{A}^{-1}\mathbf{b}$ and thereby of $c(\mathbf{x})$. However, from the complexity statements in Eqs.,\eqref{eq:runtime_nograd}, \eqref{eq:runtime_grad}, and \eqref{eq:gate_QuSO}, one can see that these errors exhibit different scaling behaviors. While $1/\epsilon$ enters logarithmically, $\delta$ contributes a linear factor $1/\delta$. Consequently, the total computational cost is asymptotically dominated by the QAE subroutine. The exact requirements for $\epsilon$ and $\delta$ are highly application-specific and depend on the spectral gap of the cost Hamiltonian. In practice, they are treated as hyperparameters determined by the desired solution quality. }

Although the present implementation of the QuSO algorithm optimizes only for the minimal temperature at a single node, it can be extended to accommodate more sophisticated cost functions. These may involve combining temperatures across multiple nodes or incorporating additional cost Hamiltonians. Furthermore, the algorithm’s flexibility allows us to consider additional constraints by modifying the mixer layer $U_\mathrm{M}(\beta)$ \cite{hadfield_quantum_2019,fuchs_constrained_2022}. 

\section{Numerical Experiments}\label{sec:simulations}
To evaluate the performance of QuSO, we implemented it for the cooling system use case. The corresponding code is available in Refs. \cite{holscher_quantum_nodate, holscher_cooling_nodate}.
\subsection{QSVT for Cooling System Simulation}
In our first simulation, we run the QSVT-based linear system solver $L$ from Eq.\,\eqref{eq:L} to evaluate the battery temperature $\tilde{T}_0$. Accordingly, the cost function is defined as 
\begin{equation}
    c(\mathbf{x})=\tilde{T}_0(\mathbf{x})\cdot 2C_\mathrm{p}C_\mathrm{B}/C_\text{LCU}^2.
    \label{eq:c(x)}
\end{equation}
The system parameters used in our simulations are listed in Table\,\ref{tab:params}.
\begin{table}[h]
    \centering
    \begin{tabular}{l|l|l}
        $R_{01} = \SI{5}{\milli\kelvin\per\watt}         $ & $\dot{Q}_{0} = \SI{2}{\kilo\watt}$     & $R_{\mathrm{env}} = \SI{10}{\milli\kelvin\per\watt}$\\
        $R_{02} = R_{03} = \SI{6}{\milli\kelvin\per\watt}$ & $\dot{Q}_{1} = \SI{4}{\kilo\watt}$     &    $T_{\mathrm{env}} = \SI{293}{\kelvin}$\\
        $R_{12} = R_{13} = \SI{7}{\milli\kelvin\per\watt}$ & $\dot{Q}_{2} = \SI{-0.2}{\kilo\watt}$  & \\
        $R_{23} = \SI{8}{\milli\kelvin\per\watt}         $ & $\dot{Q}_{3} = \SI{-2}{\kilo\watt}$    & \\
    \end{tabular}
    \caption{System parameters used for our simulations.}
    \label{tab:params}
\end{table}

\noindent The thermal resistances were chosen to have a realistic order of magnitude and to satisfy ${R_{\mathrm{env}} > R_{\min}}$, reflecting the expectation that less heat is transferred to the environment than through dedicated connections within the cooling system. At the same time, we tuned the ratio $R_{\mathrm{env}}/R_{\min}$ such that the smallest singular value $\sigma_\mathrm{min}=0.05$ remains manageable. 

\begin{figure}[h]
    \centering
    \includegraphics{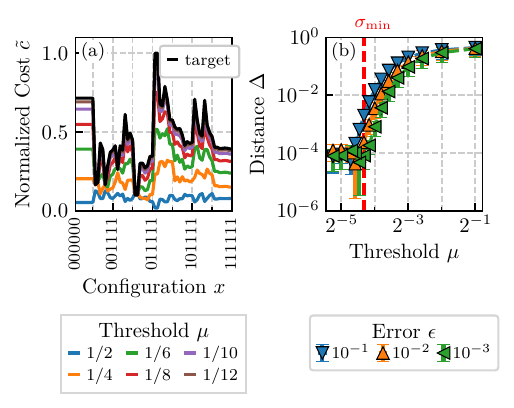}
    \caption{Simulation results for the linear system solver $L$, which incorporates the LCU block-encoding $U_\mathbf{A}$ from Eq.\,\eqref{eq:UA} and QSVT. Panel (a) presents the normalized cost $\tilde{c}$ for each configuration $x$ at various errors $\epsilon$. Panel (b) illustrates how the distance $\Delta$ (cf. Eq.\,\eqref{eq:Delta}) decreases with threshold $\mu$ and error $\epsilon$; here, the mean error is shown with standard deviation error bars.}
    \label{fig:QSVT_sim}
\end{figure}

We employ the LCU block-encoding $U_\mathbf{A}$ (cf. Eq.\,\eqref{eq:UA}) together with QSVT, utilizing the polynomial mentioned in Section\,\ref{sec:qsvt}. Its accuracy is set by the parameters $\mu$ and $\epsilon$. We ran simulations with $\mu$ ranging from $1/38$ to $1/2$, and $\epsilon$ ranging from $\SI{1e-3}{}$ to $\SI{1e-1}{}$. In order to compare the resulting cost $c(\mathbf{x})$ for all 64 configurations $x$, we normalized the cost as
\begin{equation}
    \tilde{c}(\mathbf{x})=\frac{c(\mathbf{x})}{\max\limits_{\mathbf{x}}c(\mathbf{x})}
\end{equation}
since the scaling factor $C_\mathrm{p}$ varies for different $\mu$ and $\epsilon$. Fig.\,\ref{fig:QSVT_sim}\,(a) shows the normalized cost $\tilde{c}(x)$ for several $\mu$, $\epsilon=\SI{1e-1}{}$, and all configurations $x$. 
As expected, smaller values of $\mu$ yield QSVT results that are closer to the target values $\tilde{c}_\mathrm{target}$. This trend is illustrated in Fig.\ref{fig:QSVT_sim}\,(b), where the distance
\begin{equation} 
    \Delta = \abs{\tilde{c}_\mathrm{target} - \tilde{c}} 
    \label{eq:Delta} \end{equation}
is plotted against $\mu$ for different values of $\epsilon$. Once $\mu$ is sufficiently small to resolve the smallest singular value $\sigma_\mathrm{min}$, QSVT successfully inverts the block-encoded matrix. \changemarker{Consequently, for $\mu$ below $\sigma_\mathrm{min}$, the normalized cost value $\tilde{c}$ matches its target value $\tilde{c}_\text{target}$. Note that $\Delta$ does not converge to zero for small $\mu$ but to $\Delta(\mu<\sigma_\mathrm{min})\approx 10^{-4}$. This residual error is expected, as the precision is ultimately limited by the error $\epsilon$ of the underlying QSVT polynomial.}

\subsection{QuSO for Cooling System Optimization}
Having verified that our cooling system simulation using QSVT yields correct results, we now evaluate the complete QuSO algorithm. However, the entire circuit, including the ancilla qubit registers and all subroutines within QAE, becomes computationally intensive and thus infeasible to simulate with our available hardware. \changemarker{We show qubit and gate count of an example circuit in Appendix \ref{sec:gatecount}.} To address this challenge, we employ two shortcuts only possible in statevector simulation.

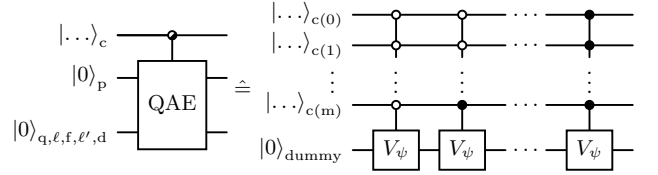
\begin{figure}[h!]
    \centering
    \resizebox{\columnwidth}{!}{
    \begin{quantikz}[row sep = 0.3cm, column sep = 0.3 cm]
    %give the wires the same vertical space as if they had H gates on them
    \lstick{$\ket{\dots}_{\texttt{c}}$} & \uctrl{} \wire[d][1]{q} & \\
    \lstick{$\ket{0}_{\texttt{p}}$} & \gate[2, disable auto height]{\mathrm{QAE}} &  \\
    \lstick{$\ket{0}_\texttt{qlfl'd}$} & & 
    \end{quantikz}$\hat{=}$\begin{quantikz}[row sep = 0.3cm, column sep = 0.3 cm]
    \lstick{$\ket{\dots}_{\texttt{c}}$} & \octrl{1}& \octrl{1} & \ \ldots\ & \ctrl{1}&\\
    \lstick{$\ket{\dots}_{\texttt{c}}$} & \octrl{1}& \octrl{1}& \ \ldots\ & \ctrl{1}& \\[-0.20cm]
    \lstick{\myvdots} \setwiretype{n}& \myvdots & \myvdots & & \myvdots &\\[-0.2cm]
    \lstick{$\ket{\dots}_{\texttt{c}}$} & \octrl{1}& \ctrl{1}& \ \ldots\ &\ctrl{1}&\\
    \lstick{$\ket{0}_{\texttt{d}}$}& \gate{V_\psi} & \gate{V_\psi} & \ \ldots\ &\gate{V_\psi} &
    \end{quantikz}
    }
    \caption{Replacement of the computationally expensive controlled QAE routine with individually controlled state preparations $V_\psi$ according to Eq.\,\eqref{eq:V_psi} for every possible cost $c(\mathbf{x})$.}
    \label{fig:dummy_qae}
\end{figure}

First, we precompute the costs $c(\mathbf{x})$ classically for all configurations $\mathbf{x}$, thereby creating a lookup table. Using this precomputed data, we define a dummy state
\begin{equation}
    \ket{\phi(\mathbf{x})}_\texttt{d} = \tilde{c}(\mathbf{x})\ket{0}_\texttt{d} + \sqrt{1-\tilde{c}(\mathbf{x})^2}\ket{1}_\texttt{d}.
\end{equation}
Now, instead of using QAE to estimate $c(\mathbf{x})$, we directly prepare the theoretical outcome of QAE \cite{nielsen_quantum_2012} to this dummy state as
\begin{align}
\begin{split}
    V_\psi\ket{0}_\texttt{p}\ket{0}_\texttt{d} &= -\frac{i}{\sqrt{2}}\sum_{j=0}^{2^k-1}e^{-i\pi\theta}\alpha_+(j)\ket{j}_\texttt{p}\ket{\psi_+}_\texttt{d}\\
    &\quad-e^{i\pi\theta}\alpha_-(j)\ket{j}_\texttt{p}\ket{\psi_-}_\texttt{d},
\end{split}
\label{eq:V_psi}
\end{align}
where $\ket{\psi_\pm}$ are given by Eq.\,\eqref{eq:psi}, $\theta$ is given by Eq.\,\eqref{eq:theta}, and $\alpha_\pm(j)$ are the QPE amplitudes given by
\begin{equation}
    \alpha_\pm(j) = \frac{1}{2^k}\sum_{l=0}^{2^k-1}e^{-i2\pi l(2^{-k}j\mp\theta)}.
\end{equation}
Thus, we replace the $\mathrm{QAE}$ and $\mathrm{QAE}^\dagger$ routines in Eq.\,\eqref{eq:costlayer} with individual state preparations $V_\psi$ that are controlled on the ${\texttt{c}\text{-register}}$ for every configuration $\mathbf{x}$, as illustrated in Fig.\,\ref{fig:dummy_qae}.

Second, since the costs are manually set via the lookup table, we use the normalized cost $\tilde{c}(x)$, reducing the required number of phase qubits $k$ due to its larger magnitude compared to the original cost $c(\mathbf{x})$ defined in Eq.\,\eqref{eq:c(x)}.

\begin{figure}[h]
    \centering
    \includegraphics{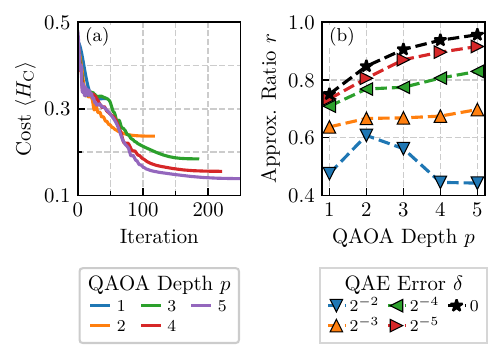}
    \caption{(a) Costs per iteration for QAOA simulations across several depths. (b) Approximation ratio $r$ for QuSO algorithm across depths $p$ and errors $\delta$.}
    \label{fig:QAOA_QuSO_cost}
\end{figure}

The best achievable performance for QuSO corresponds to standard QAOA with the cost Hamiltonian $H_\mathrm{C}$ in Eq.\,\eqref{eq:HC}. Thus, we benchmark QuSO results for different errors $\delta$ against QAOA simulations. For these simulations, the QAOA circuit parameters were initialized to $\gamma_i=\beta_i=0.5$, and we varied the circuit depth from $p=1$ to $p=5$. For minimization, we used a gradient-descent optimizer with momentum, which stopped once the cost improvement fell below a threshold of $\SI{1e-5}{}$. The resulting costs per iteration are shown in Fig.\,\ref{fig:QAOA_QuSO_cost}\,(a).
\begin{figure*}[ht!]
    \centering
    \includegraphics{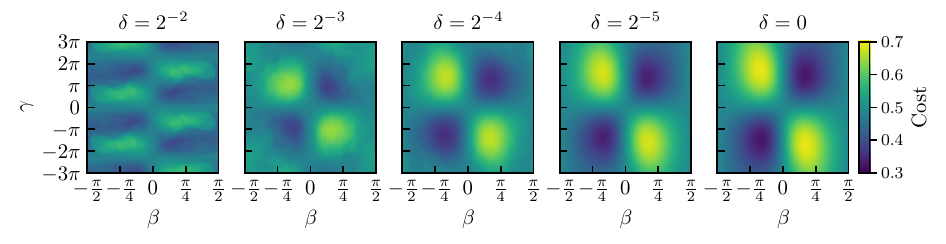}
    \caption{Cost landscapes of the QuSO cost function at depth $p = 1$ for various accuracy levels $\delta$.}
    \label{fig:cost_landscapes}
\end{figure*}
\begin{figure*}[ht!]
    \centering
    \includegraphics{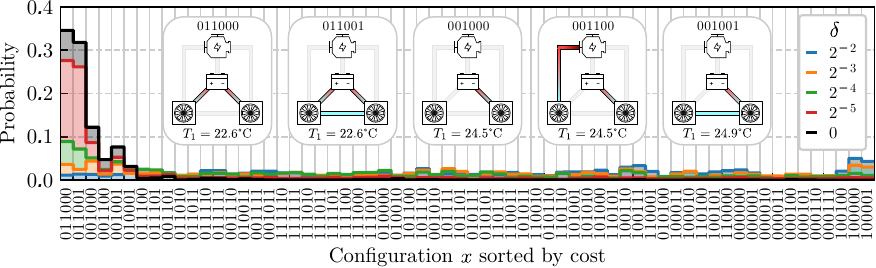}
    \caption{Probability distributions obtained after optimization with QuSO at circuit depth $p=5$ for various errors $\delta$. Bitstrings are sorted by ascending cost (optimal solutions on the left). Insets depict the cooling system configurations corresponding to the five optimal bitstrings identified by the algorithm.}
    \label{fig:QuSO_probs}
\end{figure*}

A typical performance metric in QAOA studies is the approximation ratio \cite{farhi_quantum_2014}
\begin{equation}
    r = \frac{c_\mathrm{max}-c_\mathrm{QAOA}}{c_\mathrm{max}-c_\mathrm{min}},
\end{equation}

where $c_\mathrm{QAOA}$ is the minimum cost obtained by QAOA, and $c_\mathrm{min}$ or $c_\mathrm{max}$ is the minimum or maximum cost. Consequently, $r=1$ indicates that QAOA has found the optimal solution. Fig.\,\ref{fig:QAOA_QuSO_cost}\,(b) shows the approximation ratio of QuSO at various depths $p$ and errors $\delta$. For $\delta>0$, we evaluated QuSO circuits using the optimal variational parameters $(\mathbf{\gamma}_\mathrm{opt}, \mathbf{\beta}_\mathrm{opt})$ obtained from standard QAOA simulations, whereas the results for $\delta=0$ correspond to QAOA simulations. As expected, the approximation ratio of QAOA increases with increasing depth approaching ${r=1}$ asymptotically. While the QuSO results with $\delta=2^{-2}$ show a completely different behavior, we can restore the QAOA-like behavior by further reducing $\delta$.

Similar trends emerge in the $p=1$ cost landscape plots in Fig.\,\ref{fig:cost_landscapes}, visualizing the dependence of the objective function on the parameters $\gamma$ and $\beta$. While the cost landscape with ${\delta=2^{-2}}$ differs significantly from the ideal QAOA landscape ($\delta=0$), the cost landscapes gradually converge toward the ideal behavior as $\delta$ decreases.

Finally, Fig.\,\ref{fig:QuSO_probs} illustrates the probability distributions after applying QuSO with depth $p=5$. The bitstrings on the horizontal axis are ordered according to their respective costs, with optimal solutions positioned on the left and suboptimal solutions on the right. The QAOA circuit ($\delta=0$) effectively amplifies the optimal amplitudes on the left. As observed previously, the QuSO results approach the QAOA result as $\delta$ decreases. Interestingly, at higher inaccuracies ($\delta=2^{-2}$ and ${\delta=2^{-3}}$), the probabilities of suboptimal solutions are also slightly amplified. This undesired amplification occurs due to the periodicity introduced by encoding the costs into the phase register in binary form, placing minimal cost values directly next to maximal ones. Fortunately, this effect is negligible once $\delta$ is sufficiently small.

Additionally, Fig.\,\ref{fig:QuSO_probs} shows the cooling system configurations corresponding to the five optimal bitstrings identified by the algorithm. Because our simplified cost function focuses solely on minimizing the battery temperature $T_0$, it neglects thermal interactions in other parts of the system. As a result, in the optimal cooling system only the battery is connected to both coolers, and the engine, which produces most of the heat, is isolated. Although the algorithm successfully minimized the objective function, this is of course not a practically valid solution. Hence, realistic systems require more sophisticated cost functions.

\section{Discussion}\label{sec:discussion}
Although our proof-of-concept illustrates the capabilities of QuSO, it is important to emphasize that the considered scenario is a highly simplified toy problem. Specifically, if the primary objective were to evaluate and minimize a single component of the solution vector (e.g., the battery temperature), solving just one equation would suffice instead of an entire system of equations. \changemarker{In this specific instance, we do not expect a practical quantum advantage due to the simple monotonic nature of the cost function (where increasing cooling power trivially leads to lower temperatures). However, as previously mentioned, realistic applications would require more sophisticated physics models and cost functions that consider multiple or even all entries in the solution vector. For instance, including competing objectives, such as minimizing battery temperature, ensuring a comfortable cabin temperature, and minimizing total power consumption, would create a frustrated energy landscape populated with numerous local minima. While classical gradient-based methods are prone to stagnation in such landscapes, quantum heuristics like QAOA could offer a distinct advantage in navigating the complex solution space.} \changemarker{Furthermore, our current complexity analysis benefits from the discrete graph model where the condition number $\kappa$ is bounded by fixed resistance ratios. In realistic continuum simulations (e.g., using the Finite Element Method) that consider thermal convection and radiation, mesh refinement typically causes the condition number to grow polynomially with the system size \cite{saad_iterative_2003}, which would introduce an additional polynomial overhead to the linear solver complexity.} 

It is helpful to conceptually separate QuSO into a simulation and an optimization task. On the simulation side, we show that our QAE routine encodes the phase $\theta(\mathbf{x})$ (and hence the cost $c(\mathbf{x})$) across a superposition over all $2^m$ configurations $\mathbf{x}$ with gate complexity $\propto N\log N$. Here, we neglected the precision and other problem parameters shown in Table\,\ref{tab:complexities}. While this might seem impressive, it does not give an advantage when evaluating only a single configuration, since the state collapses upon measurement. For comparison, classical linear system solvers have algorithmic complexity $\mathcal{O}\!\big(N\sqrt{\kappa}\,\log(1/\epsilon)\big)$ for the sparse case or $\mathcal{O}(N^3)$ for the dense case \cite{golub_matrix_2013}. 

On the optimization side, having $c(\mathbf{x})$ coherently encoded across a superposition of $\ket{\mathbf{x}}_\texttt{c}$ states can be used to accelerate the search over $\mathbf{x}$ to find the optimal configuration. As we currently use QAOA for that, any benefit is problem dependent and is influenced by the optimizer, initial parameters, and hyperparameters. Thus, we leave the depth $p$ and the number of classical iterations $n_{\mathrm{iter}}$ in the QuSO complexity statement. The scaling of these quantities with $N$, $\kappa$, and precision is unknown and would require its own parameter study. This is out of scope of the current work, and benchmarking becomes meaningful for larger, more realistic problems where objectives and constraints can be matched to classical approaches.

There are several promising routes for future work. On the simulation side, generalized QSVT \cite{motlagh_generalized_2024,sunderhauf_generalized_2023} could reduce circuit depth by permitting general single-qubit rotations rather than $Z$-only phases. For the outer optimization task, one may embed the cost oracle into Grover's search algorithm \cite{grover_fast_1996}, potentially providing quadratic improvements for unstructured search, as shown for topology optimization in Ref.\,\cite{holscher_end--end_2025}. Alternatively, Decoded Quantum Interferometry (DQI) \cite{jordan_optimization_2025} would also be an interesting candidate to replace QAOA. \changemarker{In general, we note that the choice of QAOA as the optimization routine is subject to ongoing research regarding its practical advantage due to limitations such as barren plateaus \cite{larocca_barren_2025, schumann_emergence_2024}. This is particularly relevant for generic higher-order constraint satisfaction problems, where recent benchmarks suggest that QAOA requires very large circuit depths ($p\gg1$) to outperform classical heuristics like the mean-field approximate optimization algorithm \cite{muller_limitations_2025}.}

\section{Conclusion}\label{sec:conclusion}
We successfully demonstrated the QuSO algorithm on a simplified toy model of a cooling system inspired by automotive thermal management. By integrating quantum algorithms, including QAOA, LCU, QSVT, QAE, and QPA, we provided an end-to-end quantum algorithm capable of simulating and optimizing the cooling system.

In conclusion, our demonstration highlights both the potential and limitations of QuSO. By explicitly analyzing the complexities of each quantum subroutine for a concrete example, this work sets the foundation for future research in quantum algorithms for classical numerical simulations and simulation-based optimization.

% \section*{Declarations}
\section*{Code Availability}
All code used to generate the figures and simulations in this article is available at \cite{holscher_cooling_nodate, holscher_quantum_nodate}.
% \subsection*{Authors' Contributions}
% L.H. designed the model and algorithm, implemented the numerical experiments, wrote the main manuscript text and prepared all figures. L.K. implemented quantum circuits and provided expertise on the engineering application. O.S. and T.D. contributed to the algorithmic design choices and circuit engineering. All authors discussed the results and revised the manuscript.
\section*{Acknowledgments}
We acknowledge the support of the BMW Group and thank J.~Stein for valuable insights and discussions. Additionally, we thank T.~Stollenwerk and F.~K.~Wilhelm for ongoing support and constructive feedback. L.K. is partly funded by the German Ministry for Economic Affairs and Energy (BMWE) in the project QCHALLenge under Grant 01MQ22008D.

\appendix

\section{Modeling the Cooling System}
\subsection{Derivation of the Linear System}\label{sec:ls}
As introduced in Section \ref{sec:problem}, our cooling system considers only heat conduction. Thus, the heat transfer rate between two nodes $i$ and $j$ is given by 
\begin{equation}
    \dot{Q}_{ij} = \frac{T_i-T_j}{R_{ij}}\mathrm{.}
    \label{eq:Q_ij}
\end{equation}
Exemplarily, we consider the fully connected graph of Fig.\,\ref{fig:problem}\,(b) with $N=4$ nodes. Since energy is conserved, the heat going in and out a node must be equal
\begin{align}
    \dot{Q}_0 + \frac{T_\mathrm{env} - T_0}{R_\mathrm{env}} = x_{01}\dot{Q}_{01} + x_{02}\dot{Q}_{02} + x_{03}\dot{Q}_{03}\\
    \dot{Q}_1 + \frac{T_\mathrm{env} - T_1}{R_\mathrm{env}} + x_{01}\dot{Q}_{01} =  x_{12}\dot{Q}_{12} + x_{13}\dot{Q}_{13}\\
    \dot{Q}_2 + \frac{T_\mathrm{env} - T_2}{R_\mathrm{env}} + x_{02}\dot{Q}_{02} + x_{12}\dot{Q}_{12} = x_{23}\dot{Q}_{23}\\
    \dot{Q}_3 + \frac{T_\mathrm{env} - T_3}{R_\mathrm{env}} + x_{03}\dot{Q}_{03} + x_{13}\dot{Q}_{13} + x_{23}\dot{Q}_{23}= 0\mathrm{.}
\end{align}
Here, the binary variables $x_{ij}$ determine if a connection and thus the heat flow from node $i$ to $j$ exists. In addition, every node is connected to the environment with temperature $T_\mathrm{env}$ and thermal resistance $R_\mathrm{env}$. If we insert Eq.\,\eqref{eq:Q_ij} in the four equations above and sort the terms by the temperatures, we obtain the following linear system of equations
\begin{widetext}
\begin{align}
    \begin{pmatrix}
        \frac{1}{R_\mathrm{env}} + \frac{x_{01}}{R_{01}} + \frac{x_{02}}{R_{02}} + \frac{x_{03}}{R_{03}} & -\frac{x_{01}}{R_{01}} & -\frac{x_{02}}{R_{02}} & -\frac{x_{03}}{R_{03}}\\
        -\frac{x_{01}}{R_{01}} & \frac{1}{R_\mathrm{env}} + \frac{x_{01}}{R_{01}} + \frac{x_{12}}{R_{12}} + \frac{x_{13}}{R_{13}} & -\frac{x_{12}}{R_{12}} & -\frac{x_{13}}{R_{13}}\\
        -\frac{x_{02}}{R_{02}} & -\frac{x_{12}}{R_{12}} & \frac{1}{R_\mathrm{env}} + \frac{x_{02}}{R_{02}} + \frac{x_{12}}{R_{12}} + \frac{x_{23}}{R_{23}} & -\frac{x_{23}}{R_{23}}\\
        -\frac{x_{03}}{R_{03}} & -\frac{x_{13}}{R_{13}} & -\frac{x_{23}}{R_{23}} & \frac{1}{R_\mathrm{env}} + \frac{x_{03}}{R_{03}} + \frac{x_{13}}{R_{13}} + \frac{x_{23}}{R_{23}}\\
    \end{pmatrix}
    \begin{pmatrix}
        T_0\\T_1\\T_2\\T_3
    \end{pmatrix}
    \notag&\\
    = 
    \begin{pmatrix}
        \dot{Q}_0+\frac{T_\mathrm{env}}{R_\mathrm{env}}\\
        \dot{Q}_1+\frac{T_\mathrm{env}}{R_\mathrm{env}}\\
        \dot{Q}_2+\frac{T_\mathrm{env}}{R_\mathrm{env}}\\
        \dot{Q}_3+\frac{T_\mathrm{env}}{R_\mathrm{env}}
    \end{pmatrix}&\mathrm{.}
    \label{eq:big_LS}
\end{align}
\end{widetext}
To eliminate the constant term $T_\mathrm{env}/R_\mathrm{env}$ on the right-hand side of the equation, we reformulate the problem by solving for the shifted temperatures $\tilde{T}_i = T_i - T_\mathrm{env}$ instead of $T_i$. This leads to the linear system presented in the main text (cf. Eq.\,\eqref{eq:ls}). The advantage of this transformation is that the vector $\mathbf{B}$ becomes sparse if the system does not have many external heat sources. The thermal connection to the environment ensures that the matrix never becomes singular, even if all binary variables are set to zero. 

For general systems with size $N$, the components of the linear system ${\mathbf{A}\mathbf{\tilde{T}}=\mathbf{B}}$ are given by
\begin{align}
    A_{ij} &= \begin{cases}
        \frac{1}{R_\mathrm{env}} + \sum_{(i, j)\in\mathbf{E}}\frac{x_{ij}}{R_{ij}} & \text{if } i=j,\\
        -\frac{x_{ij}}{R_{ij}} & \text{if } i<j,\\
        -\frac{x_{ji}}{R_{ji}} & \text{if } i>j.
    \end{cases}\label{eq:A}\\
    \tilde{T}_i &= T_i - T_\mathrm{env}\\
    B_i &= \dot{Q}_i
\end{align}

\subsection{Matrix Properties of the Linear System}\label{sec:matrixprops}
The matrix $\mathbf{A}$, defined by Eq.\,\eqref{eq:A}, is \textit{real} and \textit{symmetric} by construction. Furthermore, it is \textit{strictly diagonally dominant}, as shown:
\begin{align}
\sum_{j\neq i} \abs{A_{ij}} &< \abs{A_{ii}} \\
\Leftrightarrow \sum_{(i, j)\in\mathbf{E}} \frac{x_{ij}}{R_{ij}} &< \frac{1}{R_\mathrm{env}} + \sum_{(i, j)\in\mathbf{E}} \frac{x_{ij}}{R_{ij}} \\
\Leftrightarrow 0 &< \frac{1}{R_\mathrm{env}},
\end{align}
with $A_{ii} > 0$ since all resistances $R_{ij} > 0$. This property guarantees that $\mathbf{A}$ is \textit{positive definite} ($\mathbf{A} > 0$), implying that its eigenvalues $\lambda_i$ are positive and correspond directly to its singular values ($\sigma_i = \lambda_i$).  

The condition number of $\mathbf{A}$ is defined as:
\begin{equation}
\kappa = \frac{\sigma_\mathrm{max}}{\sigma_\mathrm{min}} = \frac{\lambda_\mathrm{max}}{\lambda_\mathrm{min}},
\end{equation}
where $\lambda_\mathrm{max}$ and $\lambda_\mathrm{min}$ denote the largest and smallest eigenvalues, respectively.  

To bound the eigenvalues, we apply the Gershgorin circle theorem \cite{gershgorin_uber_1931}, which gives:
\begin{equation}
\abs{\lambda_i - A_{ii}} \leq \sum_{j\neq i} \abs{A_{ij}}.
\end{equation}
This leads to:
\begin{align}
\lambda_\mathrm{min} &\geq \min_i \left(A_{ii} - \sum_{j\neq i} \abs{A_{ij}}\right), \\
\lambda_\mathrm{max} &\leq \max_i \left(A_{ii} + \sum_{j\neq i} \abs{A_{ij}}\right).
\end{align}

Substituting the matrix definition from Eq.\,\eqref{eq:A}, we find:
\begin{align}
\lambda_\mathrm{min} &\geq \frac{1}{R_\mathrm{env}}, \\
\lambda_\mathrm{max} &\leq \frac{1}{R_\mathrm{env}} + 2 \max_i \left(\sum_{(i, j)\in\mathbf{E}} \frac{x_{ij}}{R_{ij}}\right), \\
&\leq \frac{1}{R_\mathrm{env}} + 2 \frac{d_\mathrm{max}}{R_\mathrm{min}},
\end{align}
where $d_\mathrm{max}$ is the maximum degree of the graph (corresponding to the maximum number of connections of a component), and $R_\mathrm{min}$ is the smallest resistance value in the system. Consequently, the condition number satisfies:
\begin{equation}
\kappa \leq 1 + 2 d_\mathrm{max}\frac{R_\mathrm{env}}{R_\mathrm{min}}.
\end{equation}
Thus, the asymptotic scaling of the condition number is $\kappa = \mathcal{O}(d_\mathrm{max}R_\mathrm{env}/R_\mathrm{min})$. This result is valid for every matrix $\mathbf{A}(\mathbf{x})$ regardless of the configuration $x$.
\section{Agreement of Linear System and OpenModelica Solutions}\label{sec:modelica}

\begin{figure}[h!]
    \centering
    \includegraphics[width=0.9\linewidth]{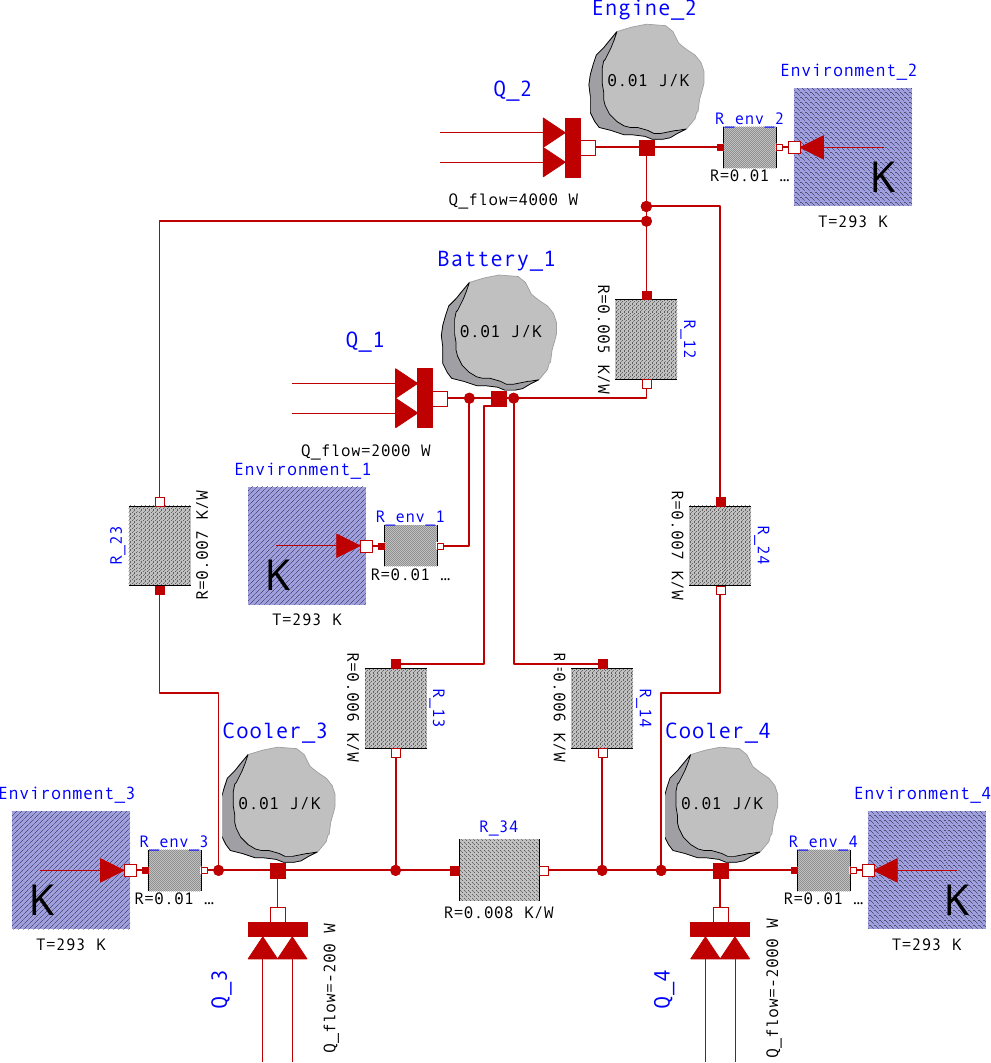}
    \caption{Visualization of the cooling system model within OpenModelica.}
    \label{fig:OMEdit}
\end{figure}

To validate our modeling of the cooling system, we compare the results of the linear system in Eq.\,\eqref{eq:big_LS} with simulations within OpenModelica \cite{fritzson_openmodelica_2020}. OpenModelica is an open-source modeling and simulation environment. Fig.\,\ref{fig:OMEdit} shows the graphical representation of the model with four nodes.

\begin{figure}[h]
    \centering
    \includegraphics{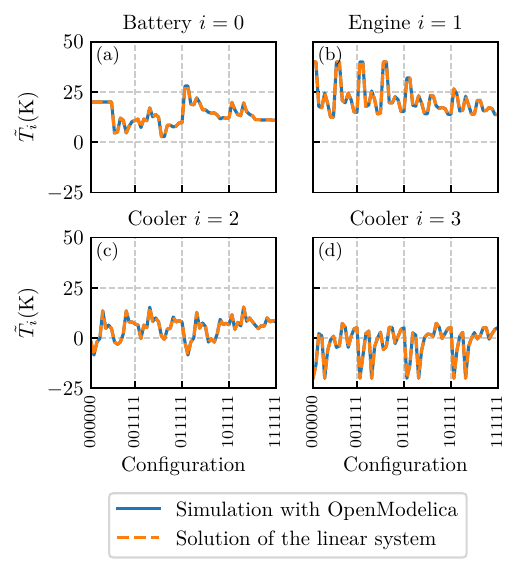}
    \caption{Comparison of simulation results using our linear system and the OpenModelica model. The temperature ${\tilde{T}_i=T_i-T_\mathrm{env}}$ is displayed for each configuration for (a) the battery, (b) the engine, and the two coolers in (c) and (d). }
    \label{fig:OM_results}
\end{figure}

\noindent Each component is represented by a heat capacitor with arbitrarily chosen thermal capacity $\SI{1}{\joule/\kelvin}$. Additionally, the heat capacitor is connected to a heat source with fixed heat transfer rate $\dot{Q}_i$ and to a thermal reservoir with temperature $T_\mathrm{env}$ via a thermal resistor with $R_\mathrm{env}$. All capacitors are connected by the corresponding thermal resistors $R_{ij}$. To compare the models, we fix the system parameters as written in Table \ref{tab:params}. The simulation results for all system configurations using both models are shown in Fig.\,\ref{fig:OM_results} and all temperatures $\tilde{T}_i$ match perfectly. Consequently, our model of a simplified cooling system is valid.

% \section{Permutation Operation $P_{ij}$}\label{sec:permutation}

\section{QuSO Cost Layer} \label{sec:qae_precision}
The QAOA cost layer is implemented in QuSO as
\begin{equation}
    U_\mathrm{C}(\gamma) = \mathrm{QAE}^\dagger \mathrm{QPA}(\gamma)\mathrm{QAE}.
\end{equation}
As defined in Eq.\,\eqref{eq:Uc}, the cost layer modifies the phase associated with each configuration $x$ in the ${\texttt{c}\text{-register}}$ by incorporating both the cost $c(\mathbf{x})$ and the variational parameter $\gamma$. As explained in the main text, this operation requires additional registers, such as the phase register p and those required for the LCU block-encoding and QSVT (q, $\texttt{l}$, f, $\texttt{l}'$, and d). In the ideal case where the p-register has enough qubits to accurately represent the phase $\theta = \arcsin(c(\mathbf{x}))/\pi$, these additional registers are uncomputed after applying the cost layer
\begin{equation}
    U_\mathrm{C}(\gamma)\ket{\mathbf{x}}_\mathrm{c}\ket{0}_\texttt{p}\ket{0} = \ket{\mathbf{x}}_\mathrm{c} \underbrace{e^{-i\gamma c(\mathbf{x})}\ket{0}_\texttt{p}\ket{0}}_{=\ket{\psi_\mathrm{ideal}}}.
    \label{eq:Uc_qae}
\end{equation}

In the following, we show that $\ket{\psi_\mathrm{ideal}}$ has the form displayed in Eq.\,\eqref{eq:Uc_qae}. The QAE routine, which includes controlled gates on the ${\texttt{c}\text{-register}}$, computes the cost $c(\mathbf{x})$ and stores the corresponding phase $\theta$ in the p-register
\begin{equation}
    \mathrm{QAE}\ket{0}_\texttt{p}\ket{0}=c_+\ket{\theta}_\texttt{p}\ket{\psi_+}+c_-\ket{1-\theta}_\texttt{p}\ket{\psi_-}.
\end{equation}
Since the Grover operator $G$ in Eq.\,\eqref{eq:grover} has two eigenvalues $\lambda_\pm=e^{\pm i2\pi\theta}$, it effectively encodes both phases $\pm\theta$ in the p-register. In the binary fraction representation, the negative phase $-\theta$ is expressed as $1-\theta$. The amplitudes $c_\pm$ have the form
\begin{equation}
    c_\pm=\mp\frac{i}{\sqrt{2}}e^{\mp i\pi\theta}.
\end{equation}
The $\mathrm{QAE}$ operation is followed by $\mathrm{QPA}(\gamma)$ to add the term $-i\gamma c(\mathbf{x})$ to the phase
\begin{align}
\begin{split}
    &\mathrm{QPA}(\gamma)\mathrm{QAE}\ket{0}_\texttt{p}\ket{0}\\
    &=c_+\underbrace{e^{-i\gamma\sin(\pi\theta)}}_{=e^{-i\gamma c(\mathbf{x})}}\ket{\theta}_\texttt{p}\ket{\psi_+}\\
    &\quad+c_-\underbrace{e^{-i\gamma\sin(\pi(1-\theta))}}_{=e^{-i\gamma c(\mathbf{x})}}\ket{1-\theta}_\texttt{p}\ket{\psi_-}\\
    &=e^{-i\gamma c(\mathbf{x})}\left(c_+\ket{\theta}_\texttt{p}\ket{\psi_+}+c_-\ket{1-\theta}_\texttt{p}\ket{\psi_-}\right).
\end{split}
\end{align}
Finally, $\mathrm{QAE}^\dagger$ uncomputes all registers besides the outer ${\texttt{c}\text{-register}}$
\begin{align}
\begin{split}
    \mathrm{QAE}^\dagger\mathrm{QPA}(\gamma)\mathrm{QAE}\ket{0}_\texttt{p}\ket{0} &= e^{-i\gamma c(\mathbf{x})}\ket{0}_\texttt{p}\ket{0}\\
    &=\ket{\psi_\mathrm{ideal}}.
\end{split}
\end{align}

In the general case where the p-register contains $k$ qubits, the $\mathrm{QAE}$ routine computes the phase $\theta$ with a finite error of $\delta=2^{-k}$. Consequently, the ancilla registers are not entirely uncomputed, and the cost layer produces
\begin{equation}
    U_\mathrm{C}(\gamma)\ket{\mathbf{x}}_\texttt{c}\ket{0}_\texttt{p}\ket{0} = \ket{\mathbf{x}}_\texttt{c}\ket{\psi_\mathrm{general}}.
\end{equation}
We now derive the exact form of $\ket{\psi_\mathrm{general}}$. Due to the error $\delta$, the $\mathrm{QAE}$ operation does not yield a single state in the p-register for $\ket{\psi_\pm}$, respectively. Instead, it creates a superposition of states
\begin{equation}
    \mathrm{QAE}\ket{0}_\texttt{p}\ket{0} = \sum_{j=0}^{2^k-1}c_+\alpha_+(j)\ket{j}_\texttt{p}\ket{\psi_+} + c_-\alpha_-(j)\ket{j}_\texttt{p}\ket{\psi_-}.
\end{equation}
Here, $\alpha_\pm(j)$ are the QPE amplitudes for $\pm\theta$ and are given by
\begin{equation}
    \alpha_\pm(j) = \frac{1}{2^k}\sum_{l=0}^{2^k-1}e^{-i2\pi l(2^{-k}j\mp\theta)}.
\end{equation}
The subsequent $\mathrm{QPA}(\gamma)$ operation yields
\begin{align}
\begin{split}
    &\mathrm{QPA}(\gamma)\mathrm{QAE}\ket{0}_\texttt{p}\ket{0}\\
    &=\sum_{j=0}^{2^k-1}c_+\alpha_+(j)e^{-i\gamma\sin(\pi 2^{-k}j)}\ket{j}_\texttt{p}\ket{\psi_+} \\
    &\quad+ c_-\alpha_-(j)e^{-i\gamma\sin(\pi 2^{-k}j)}\ket{j}_\texttt{p}\ket{\psi_-}\\
    &= e^{-i\gamma c(\mathbf{x})}\left[\sum_{j=0}^{2^k-1}c_+\alpha_+(j)\ket{j}_\texttt{p}\ket{\psi_+}+c_-\alpha_-(j)\ket{j}_\texttt{p}\ket{\psi_-}\right] \\
    &\quad+ \sum_{j=0}^{2^k-1}c_+\beta(j)\alpha_+(j)\ket{j}_\texttt{p}\ket{\psi_+}+c_-\beta(j)\alpha_-(j)\ket{j}_\texttt{p}\ket{\psi_-},
\end{split}
\end{align}
where we have introduced the term 
\begin{equation}
    \beta(j)=e^{-i\gamma\sin(\pi 2^{-k}j)}-e^{-i\gamma\sin(\pi\theta)}.
\end{equation}
The uncomputing step $\mathrm{QAE}^\dagger$ leads to
\begin{align}
\begin{split}
    &\mathrm{QAE}^\dagger\mathrm{QPA}(\gamma)\mathrm{QAE}\ket{0}_\texttt{p}\ket{0} \\
    &= e^{-i\gamma c(\mathbf{x})}\ket{0}_\texttt{p}\ket{0}\\
    &\quad+ \mathrm{QAE}^\dagger\sum_{j=0}^{2^k-1}c_+\beta(j)\alpha_+(j)\ket{j}_\texttt{p}\ket{\psi_+}\\
    &\quad+c_-\beta(j)\alpha_-(j)\ket{j}_\texttt{p}\ket{\psi_-}\\
    &=\ket{\psi_\mathrm{general}}.
\end{split}
\end{align}
We can use this expression to find an upper bound for the squared distance between the ideal and general outcome
\begin{align}
\begin{split}
    &\abs{\ket{\psi_\mathrm{ideal}}-\ket{\psi_\mathrm{general}}}^2 \\
    &= \abs{\sum_{j=0}^{2^k-1}c_+\beta(j)\alpha_+(j)\ket{j}_\texttt{p}\ket{\psi_+}+c_-\beta(j)\alpha_-(j)\ket{j}_\texttt{p}\ket{\psi_-}}^2\\
    &= \frac{1}{2}\sum_{j=0}^{2^k-1}\abs{\beta(j)}^2\abs{\alpha_+(j)}^2+\abs{\beta(j)}^2\abs{\alpha_-(j)}^2\\
\end{split}
\label{eq:dist}
\end{align}
Similarly as in Ref.\,\cite{nielsen_quantum_2012} (p.\,224), we can bound $\abs{\alpha_+(j)}^2$ using the geometric series and the inequalities $\abs{1-e^{i\varphi_1}}\leq2$ and $\abs{1-e^{i\varphi_2}}\geq2\abs{\varphi_2}/\pi$ for $-\pi\leq\varphi_2\leq\pi$:
\begin{align}
\begin{split}
    \abs{\alpha_+(j)}^2&=\abs{\frac{1}{2^k}\sum_{l=0}^{2^k-1}e^{-i2\pi l(2^{-k}j-\theta)}}^2\\
    &= \abs{\frac{1}{2^k}\frac{1-e^{-i2\pi(2^{-k}j-\theta)2^k}}{1-e^{-i2\pi(2^{-k}j-\theta)}}}^2\\
    &\leq\frac{1}{2^{2k}}\frac{4}{\abs{1-e^{-i2\pi(2^{-k}j-\theta)}}^2}\\
    %&\leq\frac{1}{2^{2k}}\frac{4}{\left(\frac{2}{\pi}2\pi\left(2^{-k}j-\theta\right)\right)^2}\\
    &=\frac{1}{2^{2k+2}}\frac{1}{\left(2^{-k}j-\theta\right)^2}.
\end{split}
\label{eq:a+}
\end{align}
The same procedure leads to 
\begin{equation}
    \abs{\alpha_-(j)}^2\leq\frac{1}{2^{2k+2}}\frac{1}{\left(2^{-k}j-1+\theta\right)^2}.
    \label{eq:a-}
\end{equation}
By elementary calculus, we find the bounds
\begin{align}
\begin{split}
    \abs{\beta(j)}^2 &= \abs{e^{-i\gamma\sin(\pi 2^{-k}j)}-e^{-i\gamma\sin(\pi\theta)}}^2\\
    &\leq\abs{\gamma\sin(\pi 2^{-k}j)-\gamma\sin(\pi\theta)}^2\\
    &\leq \gamma^2\pi^2\left(2^{-k}j-\theta\right)^2
\end{split}
\label{eq:b+}
\end{align}
and
\begin{align}
\begin{split}
    \abs{\beta(j)}^2 &= \abs{e^{-i\gamma\sin(\pi 2^{-k}j)}-e^{-i\gamma\sin(\pi(1-\theta))}}^2\\
    &\leq \gamma^2\pi^2\left(2^{-k}j-1+\theta\right)^2.
\end{split}
\label{eq:b-}
\end{align}
Combining all inequalities from Eqs.\,\eqref{eq:a+}, \eqref{eq:a-}, \eqref{eq:b+}, \eqref{eq:b-} and our expression in Eq.\,\eqref{eq:dist}, we obtain
\begin{align}
\begin{split}
    &\abs{\ket{\psi_\mathrm{ideal}}-\ket{\psi_\mathrm{general}}}^2 \\
    &= \frac{1}{2}\sum_{j=0}^{2^k-1}\abs{\beta(j)}^2\abs{\alpha_+(j)}^2+\abs{\beta(j)}^2\abs{\alpha_-(j)}^2\\
    &\leq\frac{1}{2}\sum_{j=0}^{2^k-1}\gamma^2\pi^2\left(2^{-k}j-\theta\right)^2\frac{1}{2^{2k+2}}\frac{1}{\left(2^{-k}j-\theta\right)^2}\\
    &\quad+\gamma^2\pi^2\left(2^{-k}j-1+\theta\right)^2\frac{1}{2^{2k+2}}\frac{1}{\left(2^{-k}j-1+\theta\right)^2}\\
    &=\frac{\gamma^2\pi^2}{2^{k+2}}.
\end{split}
\end{align}
Consequently, we have shown that the squared distance between the ideal state and the general state after a single cost layer scales as
\begin{equation}
    \abs{\ket{\psi_\mathrm{ideal}}-\ket{\psi_\mathrm{general}}}^2=\mathcal{O}(2^{-k})=\mathcal{O}(\delta).
\end{equation}

\changemarker{ 
\section{Qubit and Gate Count Analysis} \label{sec:gatecount} 
To illustrate why an end-to-end simulation of the QuSO algorithm was infeasible, we analyze the resource requirements for the four-node example discussed in the main text. Using our PennyLane implementation, we constructed a minimal verification circuit with the following reduced parameters: a threshold $\mu=0.5$ (resulting in a QSVT polynomial degree of 21), a QAOA depth of $p=1$, and a QAE phase register size of only $m=2$ qubits.}

\changemarker{Even with these minimal parameters, the PennyLane \texttt{specs()} function reports a circuit width of 16 qubits and a depth of 1,856,097 gates. While a 16-qubit statevector fits easily within the memory of standard workstations, the execution time is dominated by the gate count.}

\changemarker{Crucially, a scientifically meaningful simulation would require significantly higher precision. To achieve a cost function readout error of $\delta\approx10^{-3}$, the QAE phase register size must be increased to approximately $m\approx10$. Since the standard QAE algorithm requires $2^m$ applications of the Grover oracle, increasing $m$ from 2 to 10 introduces a factor of $2^8=256$ overhead. This would result in a circuit depth approaching $5\times10^8$ gates per shot. Given that the outer classical optimizer requires thousands of shots to converge, the total runtime becomes prohibitive for classical simulation.}

\changemarker{For further details and circuit statistics, we refer to the Jupyter notebook $\texttt{full\_algorithm/qubit\_gate\_count.ipynb}$ in the GitHub repository \cite{holscher_quantum_nodate}. }

% \bibliography{manual_bib}% Produces the bibliography via BibTeX.
\bibliography{references}% Produces the bibliography via BibTeX.

\end{document}